\documentclass[12pt]{article}
\usepackage{mathbbol}
\def\beq{\begin{equation}}
\def\eeq{\end{equation}}
\def\beqa{\begin{eqnarray}}
\def\eeqa{\end{eqnarray}}
\def\Pexp{{\rm Pexp}}
\def\cO{{\cal{O}}}

\def\cD{{\cal{D}}}
\def\Tr{{\rm Tr}}
\def\diag{{\rm diag}}
\def\pl{{{\cal P}_\infty}}
\newcommand{\re}{\relax{\rm I\kern-.18em R}}
\newcommand{\refeq}[1]{\mbox{Eq.~(\ref{eq:#1})}}
\newcommand{\half}{{\scriptstyle{{1\over 2}}}}
\newcommand{\thalf}{{\scriptstyle{{3\over 2}}}}

\newcommand{\veps}{\varepsilon}

\newcommand{\basispl}{
   \put(-.5,-.5){\line(1,0){1}}
   \put(.5,-.5){\line(0,1){1}}
   \put(.5,.5){\line(-1,0){1}}
   \put(-.5,.5){\line(0,-1){1}}
                         }
\newcommand{\basisar}{
   \put(0,-.5){\vector(1,0){0}}
   \put(.5,0){\vector(0,1){0}}
   \put(0,.5){\vector(-1,0){0}}
   \put(-.5,0){\vector(0,-1){0}}
	              }
\newcommand{\plaq}{\setlength{\unitlength}{.5cm}\raisebox{-.2cm}{
   \begin{picture}(1.2,1.2)(-.6,-.6)
   \basispl\basisar
   \put(-.5,-.5){\circle*{.2}}
   \put(-.55,-.55){\makebox(0,0)[tr]{\footnotesize $x$}}
   \put(-.55,0){\makebox(0,0)[r]{\footnotesize $\nu$}}
   \put(0,-.55){\makebox(0,0)[t]{\footnotesize $\mu$}}
   \end{picture}}}

\newcommand{\twoplaq}{\setlength{\unitlength}{1cm}\raisebox{-.5cm}{
   \begin{picture}(1.2,1.2)(-.6,-.6)
   \basispl
   \put(-.5,-.5){\circle*{.1}}
   \put(-.5,.5){\circle*{.1}}
   \put(.5,-.5){\circle*{.1}}
   \put(.5,.5){\circle*{.1}}
   \put(0,-.5){\circle*{.1}}
   \put(0,.5){\circle*{.1}}
   \put(.5,0){\circle*{.1}}
   \put(-.5,0){\circle*{.1}}
   \put(-.25,-.5){\vector(1,0){0}}
   \put(.25,-.5){\vector(1,0){0}}
   \put(.5,-.25){\vector(0,1){0}}
   \put(.5,.25){\vector(0,1){0}}
   \put(-.25,.5){\vector(-1,0){0}}
   \put(.25,.5){\vector(-1,0){0}}
   \put(-.5,-.25){\vector(0,-1){0}}
   \put(-.5,.25){\vector(0,-1){0}}
   \put(-.55,-.55){\makebox(0,0)[tr]{\footnotesize $x$}}
   \put(-.55,0){\makebox(0,0)[r]{\footnotesize $\nu$}}
   \put(0,-.55){\makebox(0,0)[t]{\footnotesize $\mu$}}
   \end{picture}}}

\newcommand{\linkhmu}{\setlength{\unitlength}{.5cm}\raisebox{-.2cm}{
   \begin{picture}(1.2,1.2)(-.6,-.6)
   \put(.5,0){\line(-1,0){1}}
   \put(0,0){\vector(1,0){0.1}}
   \put(-.5,0){\circle*{.2}}
   \put(-.35,-.25){\makebox(0,0)[tr]{\footnotesize $x$}}
   \put(0.4,-.3){\makebox(0,0)[r]{\footnotesize $\mu$}}
   \end{picture}}}

\begin{document}

\hfill\hbox{HU-EP-04/38, ITEP-LAT/2004-15}
\vskip5mm
\begin{center}{\Large\bf Probing for Instanton Quarks with 
\boldmath{$\veps$}-Cooling}\\
[1cm] {\bf Falk Bruckmann{${}^{(a)}$}, E.-M. Ilgenfritz{${}^{(b)}$}, 
B.V. Martemyanov{${}^{(c)}$}}\\ and\\ {\bf Pierre van Baal{${}^{(a)}$}}\\[3mm]
{\em a) Instituut-Lorentz for Theoretical Physics, University of 
Leiden,\\ P.O.Box 9506, NL-2300 RA Leiden, The Netherlands}\\[3mm]
{\em b) Humboldt-Universit\"at zu Berlin, Institut f\"ur Physik,\\
Newtonstr. 15, D-12489 Berlin, Germany}\\[3mm] 
{\em c) Institute for Theoretical and Experimental Physics,\\ 
B. Cheremushkinskaya 25, 117259 Moscow, Russia}
\end{center}
\section*{Abstract}
We use $\veps$-cooling, adjusting at will the order $a^2$ corrections to the 
lattice action, to study the parameter space of instantons in the background of
non-trivial holonomy and to determine the presence and nature of constituents 
with fractional topological charge at finite and zero temperature for SU(2). 
As an additional tool, zero temperature configurations were generated from 
those at finite temperature with well-separated constituents. This is achieved 
by ``adiabatically" adjusting the anisotropic coupling used to implement 
finite temperature on a symmetric lattice. The action and topological charge 
density, as well as the Polyakov loop and chiral zero-modes are used to 
analyse these configurations. We also show how cooling histories themselves 
can reveal the presence of constituents with fractional topological charge. 
We comment on the interpretation of recent fermion zero-mode studies for
thermalized ensembles at small temperatures. 

\section{Introduction}\label{sec:intro}

In non-Abelian gauge theories, in the absence of fields in the fundamental
representation, the Polyakov loop is an order parameter for the confinement 
to deconfinement phase transition. In the deconfined phase the center 
symmetry is spontaneously broken and the Polyakov loop is concentrated 
around center values. In the confined phase, on the other hand, the 
Polyakov loop concentrates around maximally non-trivial values, for which 
the trace vanishes. It is in such confining backgrounds that instantons 
at finite temperature (also called calorons) can dissociate in constituent 
monopoles~\cite{NCal,Lee,KvB}, all typically of the same mass, proportional 
to the temperature. 

To be more precise, this background Polyakov loop is defined in the periodic 
gauge $A_\mu(\vec x,t)=A_\mu(\vec x,t+\beta)$ by its asymptotic value, also
called the holonomy,
\beq
\pl=\lim_{|\vec x|\to\infty}\Pexp(\int_0^\beta A_0(\vec x,t)dt)=g^\dagger
\exp(2\pi i\,\diag(\mu_1,\mu_2,\ldots,\mu_n))g,
\eeq
where $g$ is the gauge rotation used to diagonalize $\pl$, whose eigenvalues
$\exp(2\pi i\mu_j)$ can be ordered on the circle such that $\mu_1\leq\mu_2\leq
\ldots\leq\mu_n\leq\mu_{n+1}$, with $\mu_{n+1}\equiv 1+\mu_1$ and $\sum_{i=1}^n
\mu_i=0$. The constituent monopoles have masses given by $8\pi^2\nu_j/\beta$ 
(their cores being of size $(4\pi\nu_j)^{-1}\beta$), with $\nu_j\equiv\mu_{j+1}
-\mu_j$. These add up to $8\pi^2/\beta$, consistent with the instanton action. 
Each constituent can be seen to carry a fractional topological charge $\nu_j$. 
For higher topological charge $Q$ the solutions are characterized by $|Q|n$ 
constituents. When well-separated they are regular 't Hooft-Polyakov 
monopoles~\cite{THPo,BPS}, where $A_0$ plays in some sense the role of the 
(adjoint) Higgs field. Their spatial locations can be chosen freely. 

It is important to note that for any temperature (no matter how small) exact 
solutions exist for which the constituents are well-separated. On the other 
hand, when constituents get closer than their size, they overlap to such an 
extent that they no longer reveal themselves as individual lumps in the 
action or topological charge density. Nevertheless, one can still uncover 
the constituents through the coincidence of two of the eigenvalues of the 
Polyakov loop, similar to what is done in Abelian projection~\cite{AbPr}. 
For example, for SU(2) half the trace of the Polyakov loop is either $-1$ or 
$+1$ at these locations.  Despite the fact that the action density follows 
closely the behaviour of normal instantons, its Polyakov loop behaves 
therefore dramatically different.

At temperatures just below the deconfining transition it has been 
well-established that a reasonable fraction of the configurations can be 
described by well-separated constituents of fractional topological charge. 
This has been studied both with cooling~\cite{Ilg2} and with fermion 
zero-modes~\cite{GatS} used as a filter to analyse Monte Carlo generated 
configurations.  In the latter case, a tell-tale signal for the constituents 
is localization of the zero-mode to constituents of different magnetic charge, 
depending on a phase $e^{2\pi iz}$ introduced for the periodicity of the 
fermions in the time direction~\cite{ZM}. For SU(2) it means that periodic 
($z=0$) and anti-periodic ($z=1/2$) zero-modes are localized on constituents 
of opposite magnetic charge, whereas for SU($n$) cycling through the boundary 
conditions the zero-mode visits constituents with the $n$ different values of 
magnetic charge.\footnote{The magnetic charges are defined with respect to the 
${\rm U}^{n-1}(1)$ subgroup that leaves the non-trivial holonomy invariant. 
One of the constituents has a charge with respect to each of the U(1) 
factors so as to make the overall configuration magnetically neutral.} 
This effect seems to persist when lowering the temperature~\cite{GatP}, 
whereas in the cooling studies with constituents still visible through 
the behaviour of the Polyakov loop, they are no longer well separated,
giving rise to instanton lumps rather than dissociated constituent 
lumps~\cite{PolyL}. 

In the course of investigating these issues we used over-improved 
cooling~\cite{OvIm} to push constituents apart. In addition we developed
two new tools that may be useful in a more general context as well. 
The first one is what we will call adiabatic cooling. This makes it possible 
to start with well dissociated constituents at finite temperature and follow 
what happens when the temperature is reduced. Finite temperature for this 
purpose is implemented on a symmetric lattice with an anisotropic
coupling~\cite{Kars}, which is also easily implemented at the level of 
improved actions~\cite{Nucu}. The anisotropy can then be brought down to 
1 in small steps, after each of which the configuration is returned to a 
classical solution by ((over-)improved) cooling. The second tool developed 
involves a more detailed analysis of the cooling history from which one can 
deduce the annihilation process of fractionally charged lumps of opposite 
duality.  For SU(2), assuming the constituents have approximately equal 
action, that is half the instanton action (defined as half a unit), such 
annihilations give a change in action of one unit and no change in the 
topological charge.  This can be contrasted with the annihilation of 
instantons, where the change in action is two units and with the case 
where an instanton falls through the lattice, in which case {\em both} 
the topological charge and action change by one unit. 

{}From the dynamical point of view cooling should be used with great 
caution to extract information on the underlying topologically non-trivial 
gauge field configurations. However, the aim of this paper is to investigate 
to which extent underlying classical (i.e.~self-dual) solutions are composed 
of localized constituents. At finite temperature we wish to study in further 
detail the case of well-separated, arbitrarily placed constituents, as well 
as the effects of overlap of constituents of equal magnetic charge giving 
rise to the doughnut structures also seen in analytic 
studies~\cite{Us}.\footnote{Remarkably, as we will see, this doughnut 
structure also comes out under (much) prolonged over-improved cooling in the 
charge 1 sector when using twisted boundary conditions~\cite{THoT,MTAP}.}
But our interest here also includes the case of low temperature, in particular 
for a lattice with the same extension in the space and (imaginary) time 
directions. It has long been conjectured that constituents, so-called {\em 
instanton quarks}~\cite{InQu}, play a role in describing the instanton 
parameter space. Indeed on the torus the $4|Q|n$ dimensions of the charge $Q$ 
moduli space of SU($n$) instantons would be most naturally described in terms 
of $|Q|n$ constituent locations of objects with topological charge $1/n$. A 
periodic array of 't~Hooft's twisted instantons~\cite{THoT,ToMa} would 
explicitly realize a corner of the moduli space that can be formulated in such 
terms, see also Ref.~\cite{ToMM}. Each such fractionally charged instanton 
lives on a smaller torus with twisted boundary conditions; gauge invariant 
quantities (like the action density) are periodic, possibly up to an element 
of $Z_n$ (like for the trace of the Polyakov loop). These fractionally charged 
instantons have a fixed scale set by the size of the small torus, such that in 
this configuration the distance between constituents is of the order of their 
size. 

This paper is organized as follows. In the next section we discuss the
notion of $\veps$-cooling, where $\veps=1$ corresponds to Wilson, $\veps=0$
to improved and $\veps<0$ to over-improved cooling. We illustrate its 
principles for a charge 1 configuration with periodic boundary conditions 
and boundary conditions where we fix the holonomies. ``Adiabatic" cooling 
is introduced,  using anisotropic couplings which is also illustrated for
the case of a charge 1 configuration with periodic boundary conditions.
These studies are extended to higher charge in Sect.~\ref{sec:chtwo} and to 
the case of charge 1 with twisted boundary conditions in Sect.~\ref{sec:tbc},
both at finite and zero temperature. In Sect.~\ref{sec:plat} we show how 
isolated self-dual action density lumps leave a clear signature in the 
cooling history at finite temperature just below $T_c$, but that above 
$T_c$ and at zero temperature these signatures are absent. We end with 
a discussion on the possible interpretation of the zero-mode results 
for thermalized configurations at low temperatures. 

\section{Cooling}\label{sec:cool}

It is well-known that for the Wilson action instantons shrink under 
cooling~\cite{BeTe}, simply because of the scaling violations due to 
the discrete lattice. This can be easily corrected by using an improved 
action. Over-improvement~\cite{OvIm} was introduced to turn the effect 
around, making the instantons grow under cooling. 
\subsection{\boldmath{$\veps$}-Cooling}
In $\veps$-cooling we can simply adjust with a single parameter the residual 
``force'' that acts on the parameters of the instanton solutions (only when 
the lattice spacing goes to zero the action does not depend on the instanton 
moduli).  For this the following lattice action ($\linkhmu\equiv U_\mu(x)
\in{\rm SU}(n)$) is used~\cite{OvIm}, 
\beq
S(\veps)=
\sum_{x,\mu,\nu}\xi_\mu\xi_\nu\left\{\frac{4-\veps}{3}\,\Tr\left(1-\plaq\right)
+\frac{\veps-1}{48}\,\Tr\left(1-\twoplaq\right)\right\},\label{eq:ecool}
\eeq
where the $\xi_\mu$, for now taken to be 1, are introduced for later 
convenience. Expanding in powers of the lattice spacing $a$ one 
finds~\cite{OvIm}, 
\beq
S(\veps)=\sum_{x,\mu,\nu}a^4\Tr\left[-\frac{1}{2}F_{\mu\nu}^2(x)+\frac{\veps 
a^2}{12\xi_\mu}(\cD_\mu F_{\mu\nu}(x))^2+\cO(a^4)\right]\label{eq:epsac2}
\eeq
(note that no summation convention is implied in this formula). $S(\veps=1)$ 
corresponds to the Wilson action, see \refeq{ecool}, and the sign of the 
leading lattice artifacts is simply reversed by changing the sign of 
$\veps$. For the initial cooling it is advantageous to use $\veps>0$ and 
only switch to $\veps<0$ when slightly above the required action, to avoid 
the solution to get stuck at higher topological charges than intended. Based 
on a discretized charge 1 infinite volume continuum instanton solution one 
finds,\footnote{Assuming $\rho\ll L$, with $L$ the size of the box, so as 
not to be affected by finite volume corrections.} $S(\veps)=8 \pi^2\{1-
\frac{\veps}{5}(a/\rho)^2+\cO(a/\rho)^4\}$, verifying that under cooling 
$\rho$ will decrease for $\veps>0$ and increase for $\veps<0$. For calorons, 
when no longer $\rho \ll\beta$, the $\cO(a^2)$ correction term will also 
depend on $\beta/\rho$ but on general grounds it can be argued to be a 
monotonic function of $\rho$ (at fixed $\beta$, in an infinite spatial 
volume). Over-improved cooling can therefore be used to separate the 
constituents, as was studied at finite temperature in Ref.~\cite{MTAP}. 

\begin{figure}[htb]
\vskip4.6cm
\includegraphics{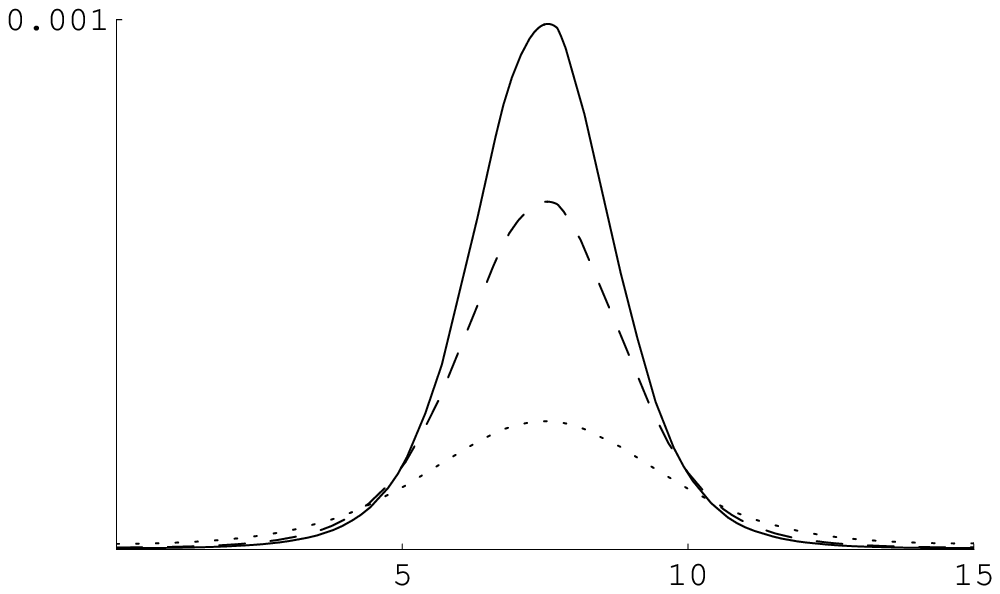}
\includegraphics{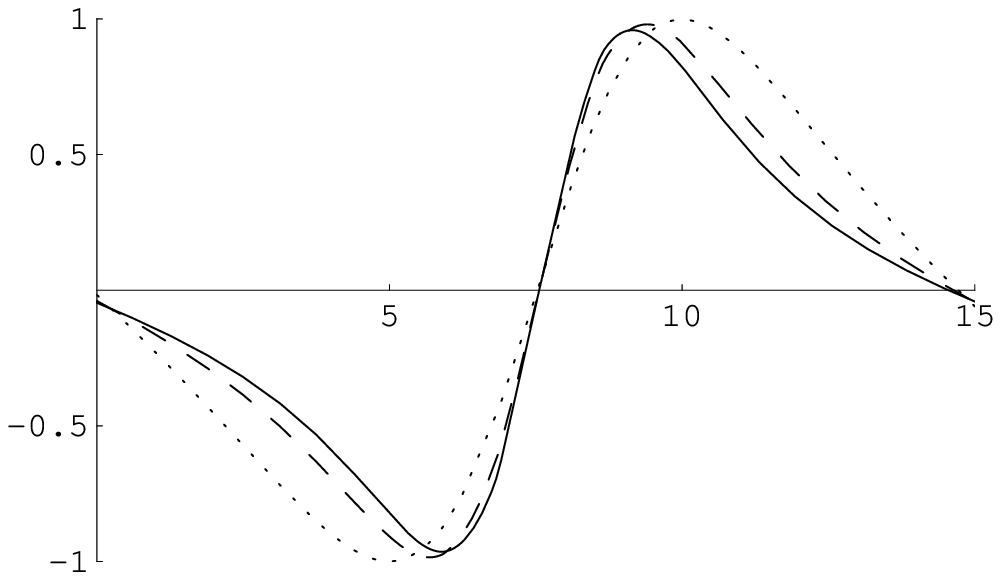}
\caption{A charge 1 configuration on a $16^4$ lattice with periodic boundary 
conditions, generated from a Monte Carlo configuration in the confined phase, 
first being cooled with $\veps=1$ to just above the one-instanton action, 
after which 500 sweeps of $\veps=0$ (full), $-1$ (dashed) and $-10$ (dotted 
curves) cooling were applied. After interpolation of the lattice data we plot 
the action density (left) and Polyakov loop (right, in one of the directions 
only) along the line connecting its extrema. From the behaviour of the 
Polyakov loop we deduce that decreasing $\veps$ pushes the constituents 
further apart.}\label{fig:fig1}
\end{figure}

The first method to study if there are localized structures at zero temperature
is to take a charge 1 configuration on a symmetric box with periodic boundary 
conditions. It had been observed in recent cooling studies~\cite{PolyL} that 
constituents did not dissociate, but could nevertheless still be unambiguously 
identified through the behaviour of the Polyakov loop reaching values of $+1$ 
and $-1$ within the single instanton action density lump, as long as the 
holonomy is non-trivial.\footnote{In a finite volume the holonomy is 
determined by averaging the trace of the Polyakov loop, which typically
agrees well with its average value in the low action density regions, as was 
used in Ref.~\cite{Ilg2}.} With over-improved cooling we can now push these 
constituents further apart and investigate whether there is a regime where 
they could reveal themselves as individual constituent lumps, as happens at 
finite temperature. There is one obstacle that makes this study somewhat 
cumbersome. For the torus without twisted boundary conditions no regular 
charge 1 self-dual solution exists, as could be proven rigorously in the 
context of the Nahm transformation~\cite{Bra}. 
Taubes had shown earlier that no obstructions exist for higher topological 
charge~\cite{Tau}. There is no problem in having configurations with 
topological charge 1, like taking an infinite volume solution whose bulk 
part fits on the torus and in the low action density region only requires 
minor modifications to adjust to the boundary conditions for the torus. But 
these configurations no longer can be exactly self-dual when their size 
remains finite. This means they shrink even when cooling with an action that 
has no lattice artifacts. With sufficient over-improvement, the obstruction 
can be counteracted as is shown on a $16^4$ lattice in Fig.~\ref{fig:fig1}.

\begin{figure}[htb]
\vskip7.8cm
\includegraphics{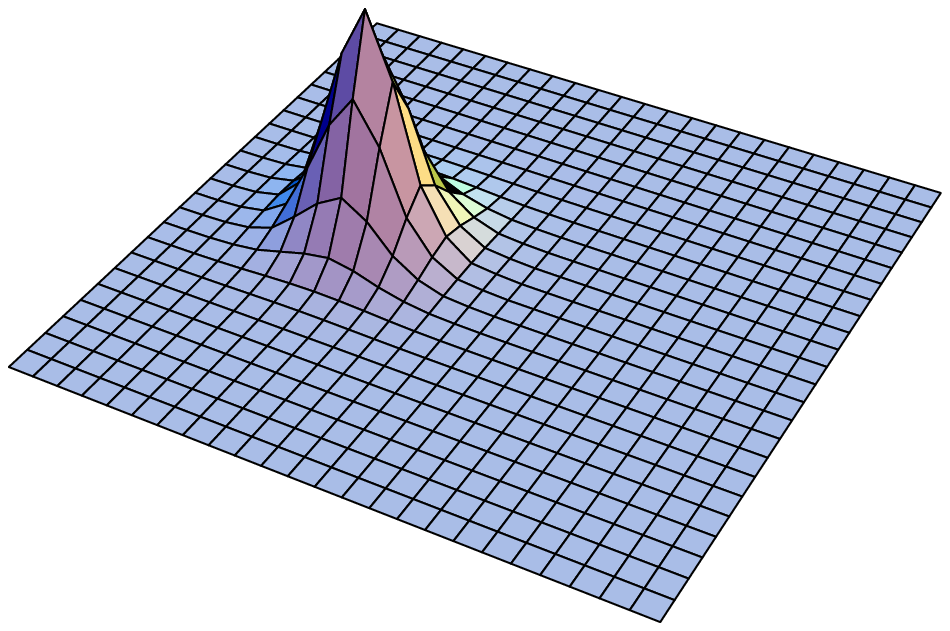}
\includegraphics{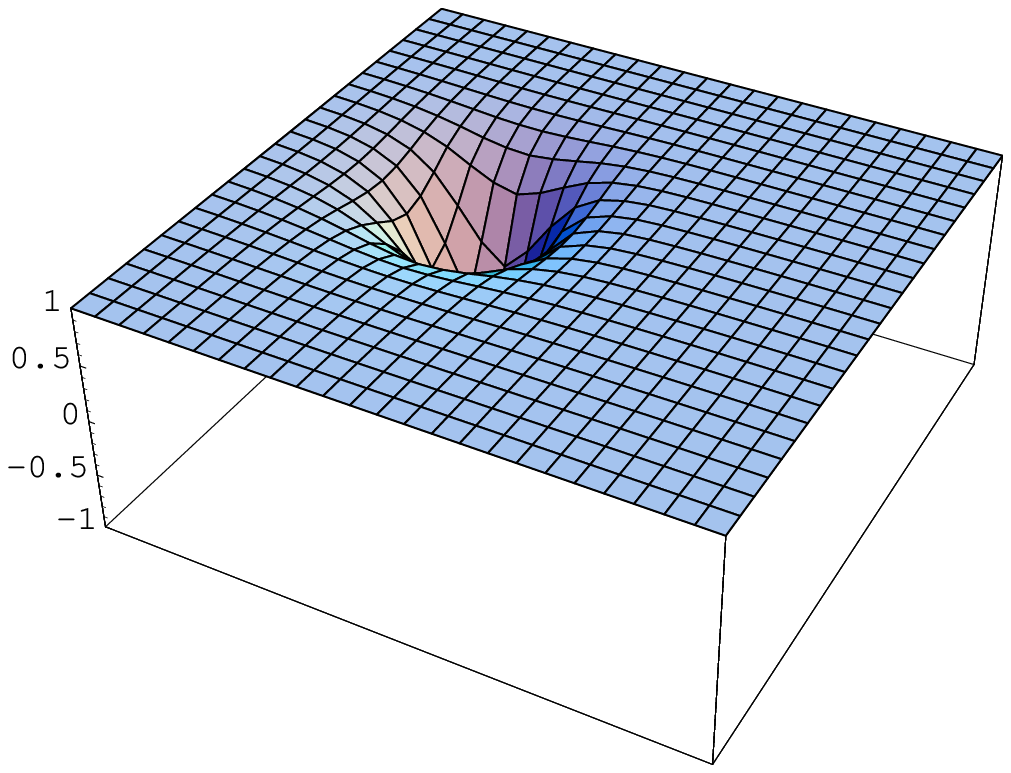}
\includegraphics{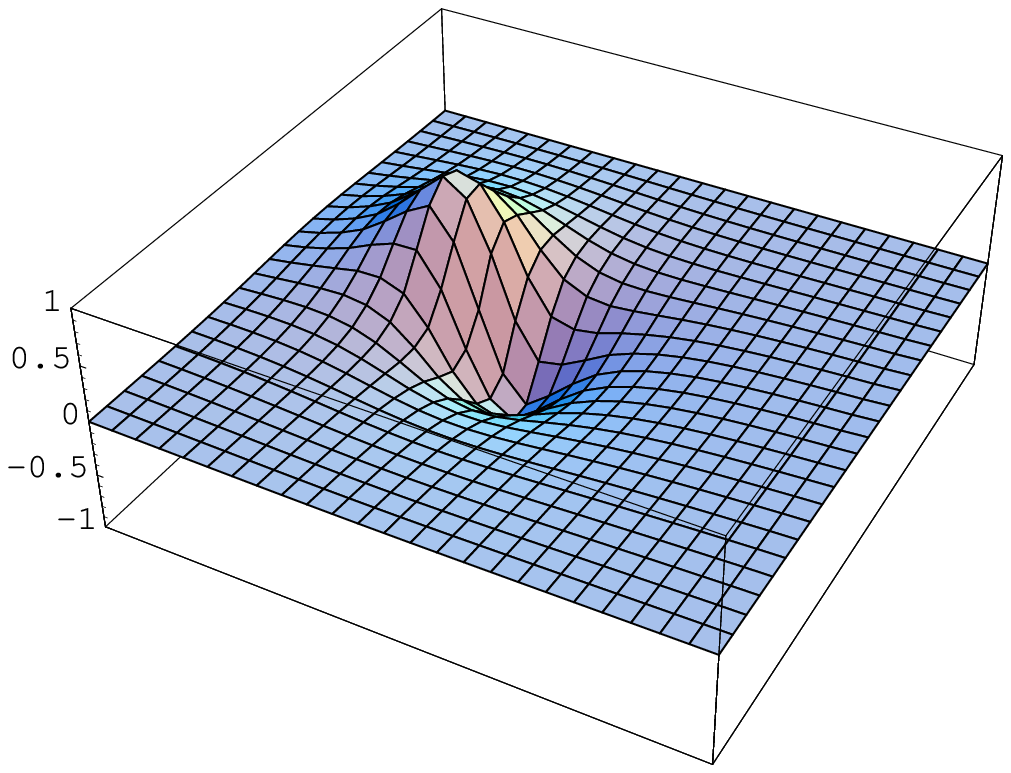}
\caption{A charge 1 configuration on a $16^4$ lattice with the holonomies 
fixed to be trivial in one direction and maximally non-trivial in the other 
three directions, generated from a random start first being cooled with $\veps
=1$ to just above the one-instanton action, after which 80 sweeps of $\veps=-1$ 
cooling were applied. We plot the Polyakov loop for two relevant directions, 
in a plane through the center of the instanton. In this plane the action 
density is shown in the middle.}\label{fig:fig2}
\end{figure}

With periodic boundary conditions we cannot keep the holonomies in the various
compact directions fixed. We anticipate on the basis of the caloron studies 
that the constituent nature comes out best in case these holonomies are 
maximally non-trivial. One sure way to enforce non-trivial holonomy, as has 
been implemented in the finite temperature cooling studies~\cite{Ilg1}, is by 
choosing appropriate fixed boundary conditions for the time-like links. On a 
symmetric box there is no preferred direction that plays the role of the 
imaginary time and Polyakov loops in all other directions are expected to 
behave similarly. Unlike at finite temperature, the holonomies in the space 
directions can now also be non-trivial. The most efficient way to fix the 
holonomy in the direction $\mu$ to $V_\mu$ is to take at $x_\nu=1$ (for $\nu
\neq\mu$) all links $U_\mu(x)$ to be independent of the remaining three 
coordinates, and equal to $\bar{U}_\mu$ such tat $V_\mu=\bar{U}_\mu^{N_\mu}$. 
In Ref.~\cite{Syn} these holonomies were shown to play a role in fixing 
instanton moduli on the torus. For the cases studied there, the holonomies 
are mapped to constituent locations under the Nahm transformation. The choice 
of holonomy indeed strongly influences the local behaviour of the Polyakov 
loop. When the holonomy is non-trivial there is a characteristic ``dipole" 
structure, but for trivial holonomy the structure is like a ``monopole", as 
illustrated in Fig.~\ref{fig:fig2}.

Due to the fixed boundary conditions one can not expect to be able to find 
exactly self-dual configurations, but we can again use over-improved cooling 
to attempt to separate constituents. The behaviour of the Polyakov loop did 
show that over-improved cooling has the desired effect, but we could not 
reach the stage where isolated action density lumps were revealed. 

\subsection{Adiabatic Cooling}\label{sec:adi}

In our search for constituents at low temperatures, we can make use of our 
knowledge at finite temperature, starting with a configuration that has well 
localized constituents. Subsequently the temperature is lowered in small 
steps, after each step applying ((over)-improved) cooling to re-adjust the 
configuration to a (near) solution. We call this process adiabatic cooling. 
Implementing this by adding a time slice to the lattice to lower the 
temperature, one has to worry {\em how} to extend the configuration to this
additional slice, and whether the (discrete) change in temperature is not 
too big a perturbation. Both of these problems are solved when using 
anisotropic couplings on a symmetric lattice to implement finite 
temperature~\cite{Kars}, since the anisotropy can be changed continuously.
It is for this reason we introduced $\xi_\mu$ in \refeq{ecool}. In this
form all aspect ratios can be changed continuously. With $a_\mu\equiv a/
\sqrt\xi_\mu$ the expansion of $S(\veps)$ in \refeq{ecool} is as given 
in \refeq{epsac2}. We fix $\prod_\mu\xi_\mu=1$, such that when approximating 
the sum over the lattice points by an integral the leading term correctly 
corresponds to $-\half\int d^4x\Tr F_{\mu\nu}^2(x)$, since the proper volume 
element of a lattice cell is $\prod_\mu a_\mu=a^4$.

For finite temperature the $\xi_\mu$ are as usual parameterized by {\em one} 
anisotropy parameter $\xi$, with $\xi_0=\xi^\thalf$ and $\xi_i=\xi^{-\half}$. 
This implies that the lattice spacing in the time direction is a factor 
$\xi$ smaller than in the space direction, $a_t=a_s/\xi$ (or $a_0=a_i/\xi$), 
such that a lattice of size $N_s^4$ with an anisotropy parameter $\xi$ is 
equivalent to a lattice of size $N_t\times N_s^3$, with $N_t=N_s/\xi$. Our 
studies for isotropic lattices at finite temperature are with a size 
$4\times 16^3$. This would therefore be equivalent to results on a lattice 
of size $16^4$ with an anisotropy parameter $\xi=4$. Reducing $\xi$ to 1
under adiabatic cooling gives results on isotropic lattices of size $16^4$,
which is the situation implied when we talk about zero temperature.

In our adiabatic cooling studies we used two methods to create the initial 
configurations at finite temperature. The simplest is to take an exact infinite 
volume and finite temperature continuum solution with the desired properties, 
naively discretized on the anisotropic lattice (by approximating the path 
ordered integral for the gauge field along the link by 30 steps of equal 
length), and performing a number of cooling sweeps to adjust it to periodic 
boundary conditions. The other method is first to use the results obtained 
from cooling on an isotropic but asymmetric lattice to get the desired 
finite temperature lattice configuration (e.g.~using over-improvement to 
separate the constituents). This configuration can then be put on the finer 
anisotropic lattice by splitting the time-like links in $\xi=N_s/N_t$ equal 
factors and for the space-like links by using a geodesic interpolation on the 
group manifold,
\beqa
U_0'({\vec x},\xi(x_0-1)+k)&=&\left[U_0({\vec x},x_0)\right]^{\frac{1}{\xi}},
\quad\qquad\qquad k=1,\ldots,\xi,\nonumber\\
U_i'({\vec x},\xi(x_0-1)+k)&=&\left[U_i({\vec x},x_0+1)U^{-1}_i({\vec x},x_0)
\right]^{\frac{k-1}{\xi}}U_i({\vec x},x_0).
\eeqa
Again, some cooling sweeps are needed to relax the configuration to a 
solution on the anisotropic lattice. Both methods work equally well to
find a starting configuration at finite temperature on the anisotropic
lattice with well-separated constituents. 

\begin{figure}[htb]
\vskip4.7cm
\includegraphics{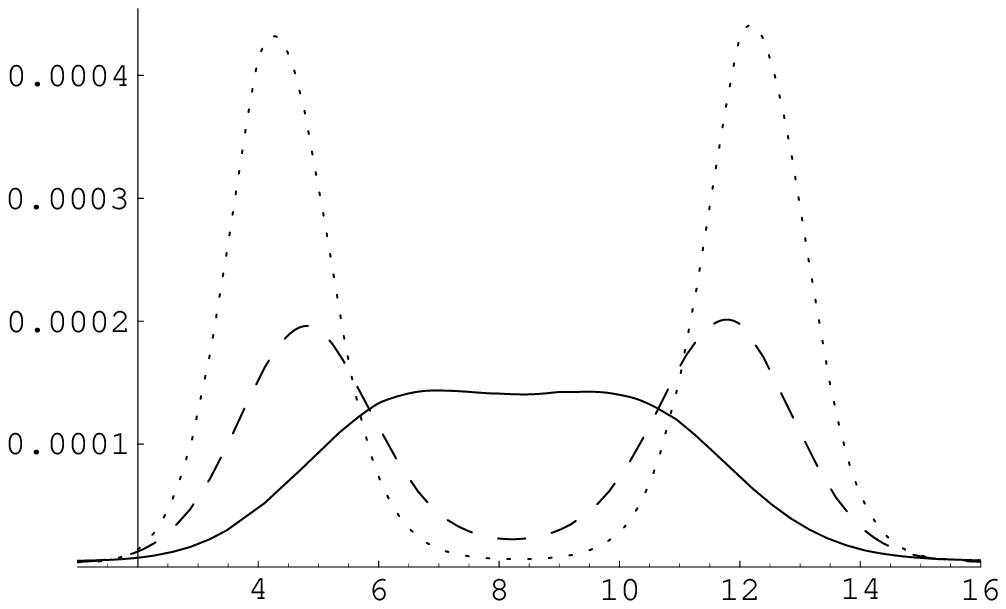}
\includegraphics{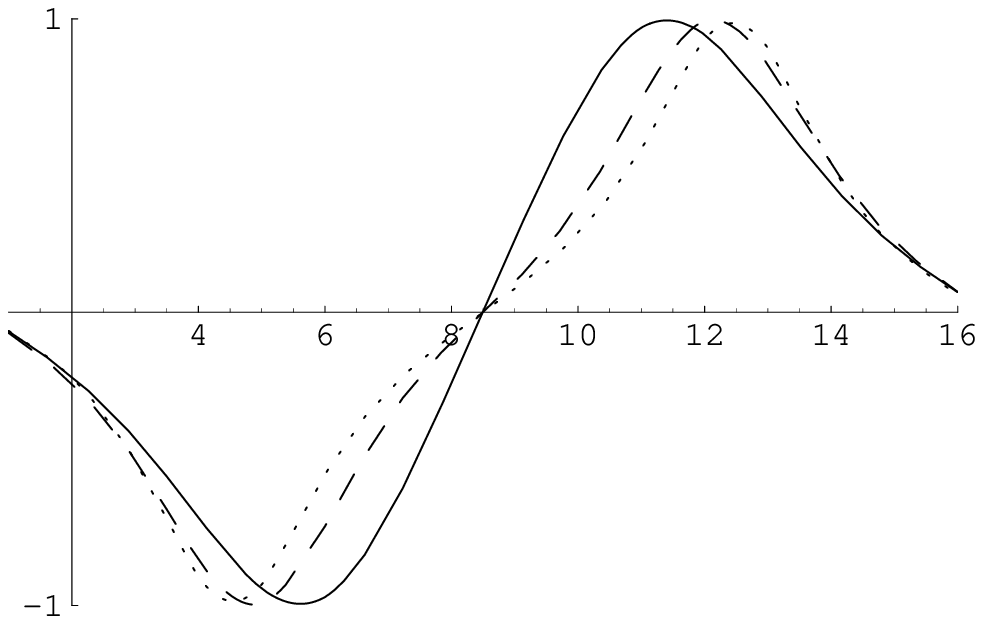}
\caption{Starting from a continuum caloron solution with well-separated
lumps, discretized on the anisotropic lattice and adjusted by 100
$\veps=-10$ cooling sweeps, we performed the adiabatic cooling by reducing
$\xi$ from 4 to 1, through $\xi=2\sqrt 2$, 2 and $\sqrt 2$, applying between
each of the 4 steps 100 $\veps=-10$ cooling sweeps. Shown is on the left 
the action density and on the right the Polyakov loop in the time direction 
along a line through the constituent locations. The dotted, dashed and full 
curves are for $\xi=4$, 2 and 1, respectively.}\label{fig:fig3}
\end{figure}
For charge 1, discretizing the infinite volume caloron solution is more 
convenient in making finite temperature configurations with well separated 
constituents due to the charge 1 obstruction on a torus. This was used to 
generate Fig.~\ref{fig:fig3}. We observe that each of the two separate 
lumps is growing in accordance of what would happen in the infinite volume 
when lowering the temperature. Increasing overlap leads to increasing 
non-static behaviour, but before the constituents become localized in 
all four directions, they have formed a single instanton lump. 

As before, the behaviour of the Polyakov loop still allows us to identify 
the constituent locations. The fact that these come a little bit closer under 
the process of adiabatic cooling is mainly due to the obstruction for having 
exact solutions of charge 1 on a torus. There are two ways to avoid this 
finite volume obstruction, either by using higher topological charge or by 
the use of twisted boundary conditions, discussed in the next two sections.

\section{Higher Charge Configurations}\label{sec:chtwo}

\begin{figure}[htb]
\vskip7.5cm
%\special{psfile=fig4a.eps voffset=-5 hoffset=50 vscale=75.0 hscale=75.0}
\includegraphics{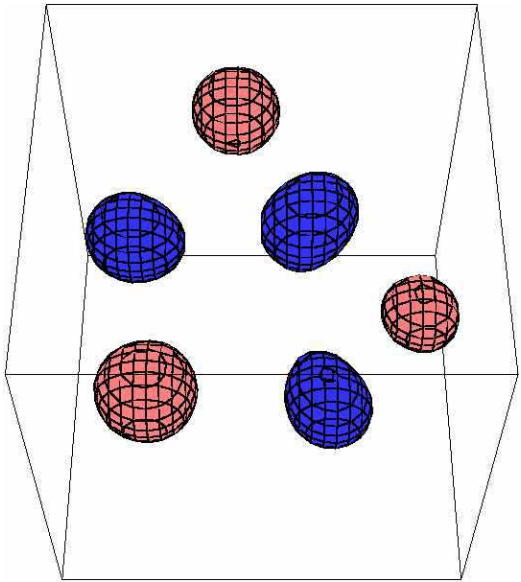}
\caption{Example of a charge 3 caloron solution with $\half\Tr\pl=-0.126$ 
on a $4\times 16^3$ lattice obtained from $\veps=-1$ cooling. Shown are 
the surfaces where half the trace of the Polyakov loop takes on the values 
$0.5$ (light, red) and $-0.65$ (dark, blue), corresponding respectively to the 
constituent monopoles with positive and negative magnetic charge.
}\label{fig:fig4}
\end{figure}

First we study in more detail the caloron moduli space at finite temperature. 
The interest here is two-fold. In our analytic studies we have seen that well 
separated constituents become point like~\cite{Us} (i.e. spherically symmetric 
BPS monopoles~\cite{BPS}), but a full analytic understanding on the moduli 
space is not yet available. Properly manipulating $\veps$ in our cooling 
studies, configurations can be found where the constituents are well-separated
and are arbitrarily positioned. An example of a charge 3 caloron solution 
with non-trivial holonomy is shown in Fig.~\ref{fig:fig4}. One clearly 
distinguishes the 6 constituents, 3 of positive and 3 of negative magnetic 
charge.

Next we study what happens at finite temperature under extended over-improved 
cooling. Oppositely charged constituents are known to be repelled and to 
become of equal mass under this cooling~\cite{MTAP}. It is therefore natural
to expect that constituents with equal charges are attracted. We should 
emphasize here again that this force is exclusively due to the lattice 
artifacts. It offers us an opportunity to move around in the moduli space. 
It should be understood though that the control one has is limited, since 
only one parameter $\veps$ is available to manipulate all (non-trivial) 
moduli. 
\begin{figure}[htb]
\vskip9.3cm
%\special{psfile=fig5a.eps voffset=105 hoffset=0 vscale=60.0 hscale=60.0}
\includegraphics{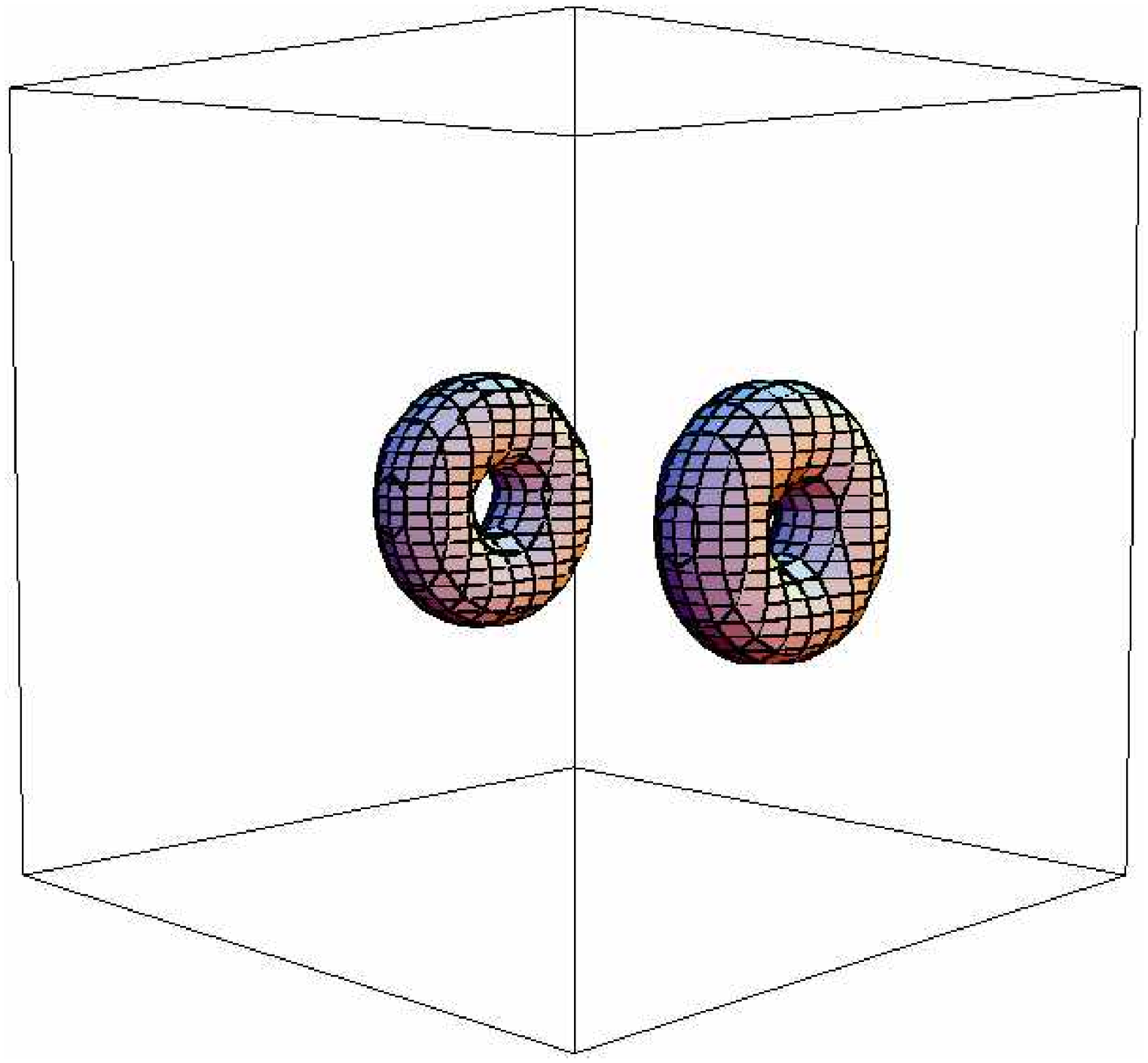}
\includegraphics{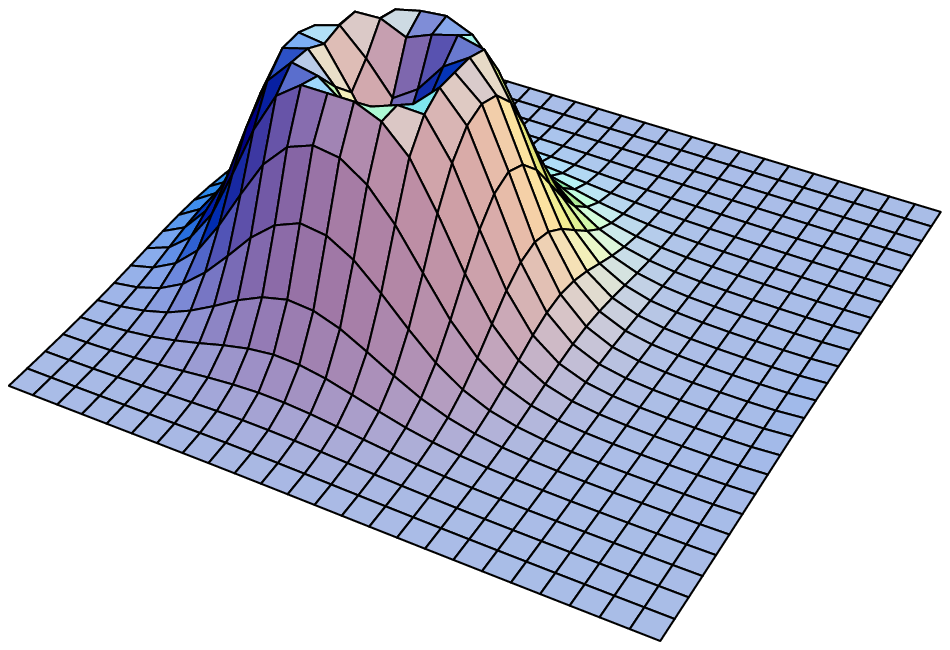}
\includegraphics{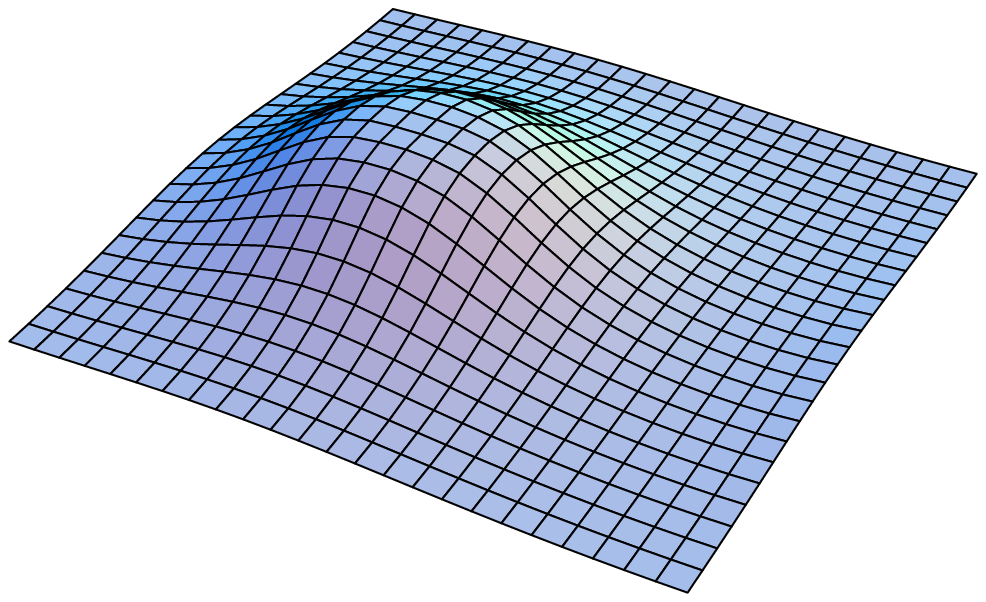}
\includegraphics{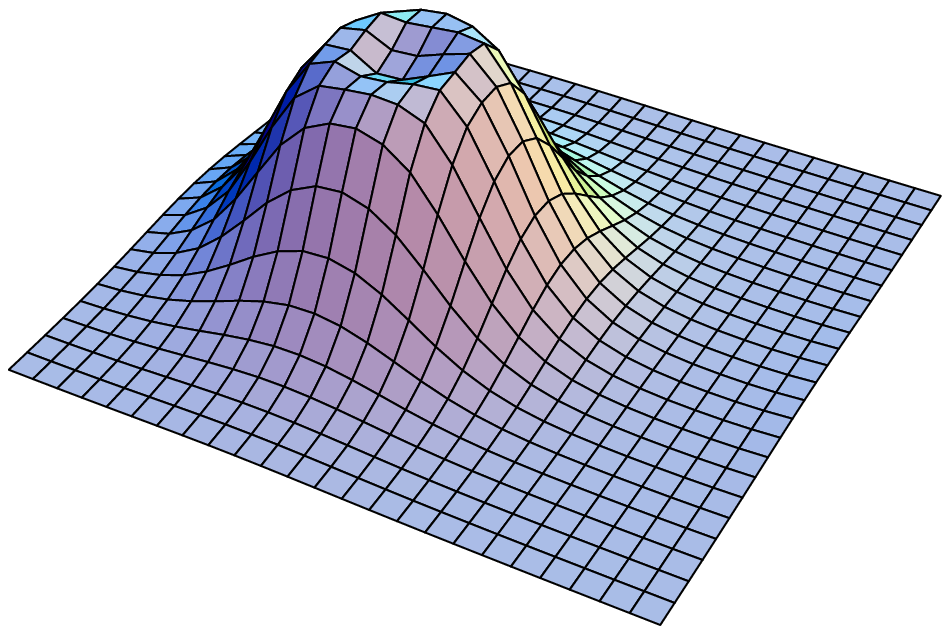}
\caption{A charge 2 caloron with $\Tr\pl=0$ on a $4\times 16^3$ lattice 
obtained from a Monte Carlo generated configuration at $4/g^2=2.2$. We first 
went down to slightly above the 2 instanton action with $\veps=1$. After that 
many thousands of $\veps=-2$ cooling sweeps (followed by 500 with $\veps=0$) 
were performed. This gives the finite volume modification of the so-called 
``rectangular" solution constructed in Ref.~\cite{Us}. Shown is a suitable 
surface of constant action density for the double doughnut structure, as well 
as (clockwise) the action density, the periodic zero-mode density and the 
Polyakov loop. The latter three are shown on a plane through the doughnut 
which supports the periodic zero-mode. The other doughnut seen by the action 
density and the anti-periodic zero-mode has the sign of the Polyakov loop 
inverted, but is not seen by the periodic zero-mode.}\label{fig:fig5}
\end{figure}
Nevertheless, this provided us with sufficient control to find 
for charge 2 that both constituents with the same charge will approach
each other. Ultimately they will be on top of each other, forming the 
doughnut structure characteristic of the axially symmetric charge 2 monopole 
solutions~\cite{MonR}. Another characteristic of these solutions is the 
double zero in the Higgs field as reflected here in the behaviour of the 
Polyakov loop. At the same time the two doughnuts, which have opposite magnetic 
charge, are repelled and will be placed as far apart as is allowed by the 
finite volume. This is illustrated in Fig.~\ref{fig:fig5}.

Next we applied adiabatic cooling to the configuration in Fig.~\ref{fig:fig5}.
Under very long over-improved cooling this configuration is actually reaching
the exact charge 2 self-dual constant curvature solution that can 
exist on a symmetric torus~\cite{THoC,PvBC}. For other aspect ratios constant 
curvature solutions exist as well, but are in general no longer self-dual 
and thus unstable~\cite{PvBC}, or at best marginally stable at non-trivial 
holonomy~\cite{PvBM,IMMP}. An expansion in the aspect ratio was performed in 
Ref.~\cite{DefC} to investigate to which type of a self-dual configuration 
these constant curvature solutions deform.  We will not discuss this here 
in greater detail, but we did see similar extended structures in case $\xi$ 
was close to 1. The analysis in Ref.~\cite{DefC} was for the self-dual constant
curvature solution of topological charge 1/2, based on suitably chosen twisted 
boundary conditions (in the 0-3 and 1-2 planes only),\footnote{To avoid any 
possible confusion we point out that the charge 1/2 building blocks mentioned 
in the introduction have twist in all the 6 possible planes, in which case 
the self-dual configuration can not be of constant curvature. By symmetry 
considerations it is localized equally in all four direction on a symmetric 
box. Its size is set by the size of this box~\cite{ToMa}.} with the sides of 
the four dimensional box satisfying $L_0L_3=L_1L_2$. To get the case we 
studied, one combines four of these boxes to a symmetric box with no twisted
boundary conditions.

\begin{figure}[htb]
\vskip4.5cm
\includegraphics{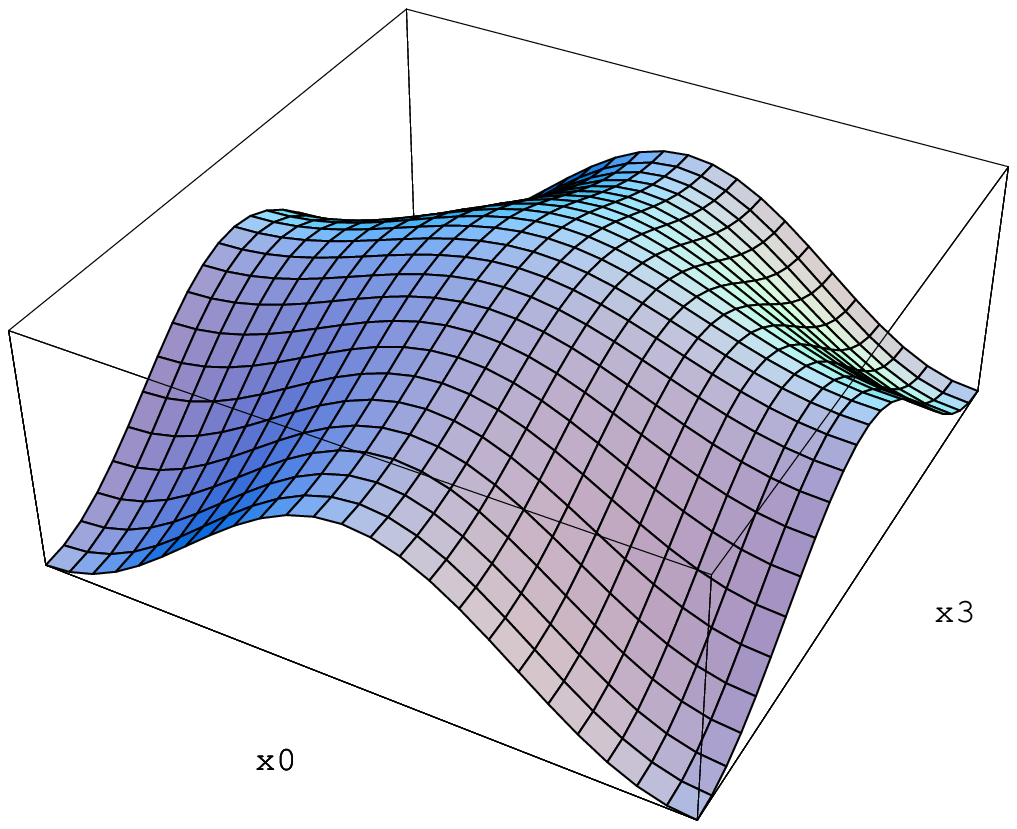}
\includegraphics{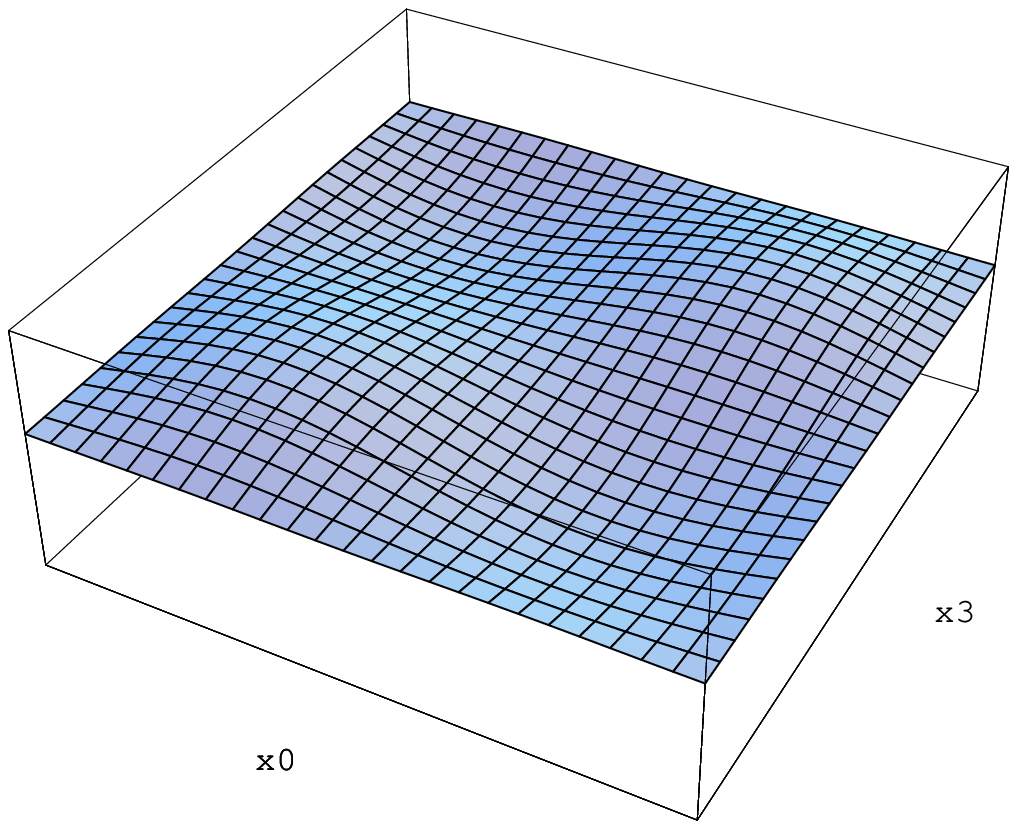}
\caption{Sum of the two exact zero-mode densities for a charge 2 constant 
curvature configuration. The flux has non-zero components in the 1-2 and 
0-3 planes. The result is plotted as a function of $x_0$ and $x_3$ for 
$x_1=x_2=z_1=z_2=0$, on the left at $z_0=z_3=0$ (the two zero-mode densities 
fall on top of each other) and on the right at $z_0=z_3=0.25$ (the two 
zero-modes densities are shifted by half a period in the $x_0$ and $x_3$ 
directions relative to each other).}
\label{fig:fig6}
\end{figure}

Although these constant curvature configurations are rather special to the 
finite volume, it is nevertheless clear what gives rise to these extended 
structures, when one attempts to separate constituents. Lowering the 
temperature their size increases. For the symmetric box this size becomes of 
the order of (half) that of the volume, as this is the only length scale in 
the system when viewing a constituent in isolation. On the other hand the 
separation between the constituents can not get bigger than the size of the 
volume. The constituents are therefore {\em bound} to overlap, and in 
general will show a single instanton peak, which increases in height with 
decreasing constituent separation. Apart from exceptional cases, built from 
periodic arrays of charge 1/2 instantons~\cite{ToMa} as discussed before, 
well-separated constituents do not reveal well-localized lumps of fractional 
topological charge, despite the fact that the underlying constituent 
description seems undeniable, as revealed by the behaviour of the Polyakov 
loop. An interesting question is now whether for these very extended 
structures the chiral fermion zero-modes still follow the underlying gluonic 
distribution. At finite temperature these zero-modes are exponentially 
localized to the cores of the constituents~\cite{Us}. At zero temperature 
there can be no exponential localization in the classical background 
field. From this point of view it is interesting to study the zero-modes 
for the self-dual charge 2 constant curvature solution. These were 
constructed before in the context of the Nahm transformation~\cite{Buck}. 
For the sum of the two zero-mode densities see Fig.~\ref{fig:fig6}.

As we can see from this figure the zero-modes do in general not have constant 
density,\footnote{Although the action density is constant, the gauge field is
not as is seen from the expression for the Polyakov loop, $\half\Tr\,\Pexp
(\int_0^{L_\mu}\!A_\mu(x)dx_\mu)=\cos(2\pi\eta^3_{\mu\nu}x_\nu/L_\nu)$, with 
$\eta^a_{\mu\nu}$ the 't Hooft tensor.} but can of course also not be 
considered to be localized. As already mentioned in the case of finite 
temperature, the zero-modes depend on the choice of boundary conditions 
for the fermions. These can be periodic up to an arbitrary phase $\exp(2\pi 
iz_\mu)$, here in each of the four directions, which is equivalent to adding 
$-2\pi iz_\mu\Eins_n/L_\mu$ to the gauge field, as is customary in formulating 
the Nahm transformation~\cite{NCal,Bra}. In the natural basis used in 
Ref.~\cite{Buck} the two zero-modes shift in opposite directions as a function 
of $z_\mu$ and happen to have the same shape. We give the sum 
of the zero-mode densities in the $0$-$3$ plane for the two cases described in 
the caption of Fig.~\ref{fig:fig6} (a slightly better ``localization" can be 
found in the $0$-$1$ plane~\cite{Marga}).

One might argue that it is not a surprise to ultimately end up in the least 
localized configuration possible for the symmetric box under adiabatic cooling, 
since our starting point was (what we believe to be) the least localized 
configuration allowed at finite temperature, as shown in Fig.~\ref{fig:fig5}.
To check that in a symmetric box the constituents indeed become as big as 
(half) the volume, we instead start at finite temperature with a well-localized 
configuration. To do this we could take an infinite volume charge 2 analytic 
solution with four well-separated constituents, that still fits sufficiently 
well into the finite volume under consideration. The most suitable 
configuration for this purpose is the so-called ``crossed" configurations 
considered in Ref.~\cite{Us}, without a net dipole moment. But rather than 
following the cumbersome procedure of putting 
this exact charge 2 solution on the lattice, we make use of the efficiency
of cooling to quickly settle down to a nearby solution. We thus take a charge 
1 caloron solution whose two constituents are separated by half a period (i.e. 
8 lattice spacings, twice the period in the imaginary time direction), and 
add to this gauge field the same solution rotated by 180 degrees and shifted 
perpendicular to its axis over half a period. As had been discussed 
extensively for the continuum in Ref.~\cite{BrvB}, this gives rise to 
would-be Dirac strings becoming visible, i.e.~carrying action density.
When simply adding two self-dual solutions, this is certainly the most 
conspicuous source for the violation of self-duality. Nevertheless, we have 
seen that cooling very quickly removes these would-be Dirac strings 
and automatically performs the exponential fine-tuning that would have been 
required in the continuum (the coarseness of the lattice in this respect 
has its advantages now). 
\begin{figure}[htb]
\vskip7.8cm
\includegraphics{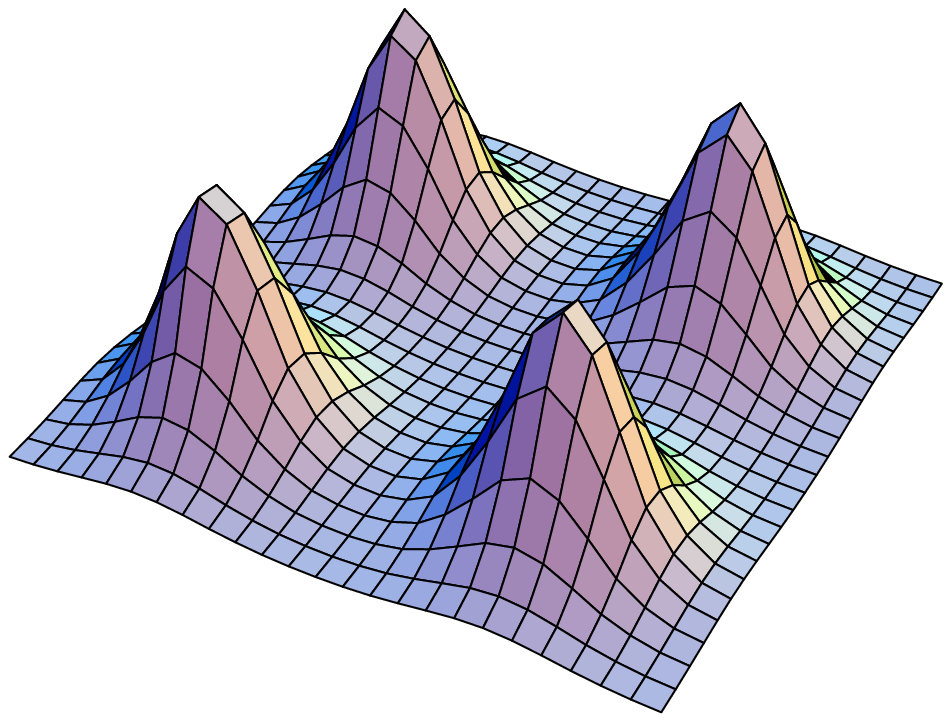}
\includegraphics{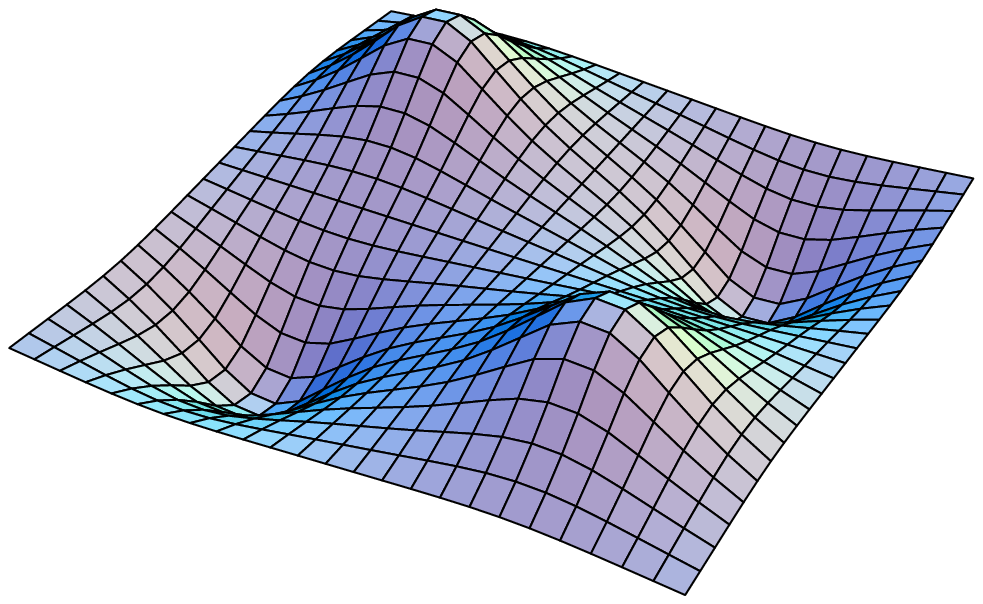}
\includegraphics{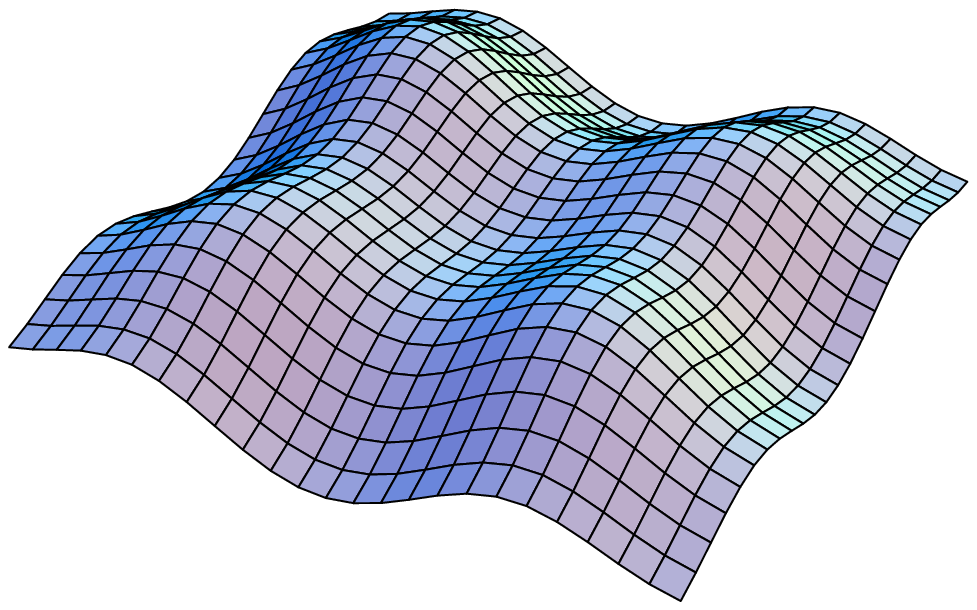}
\includegraphics{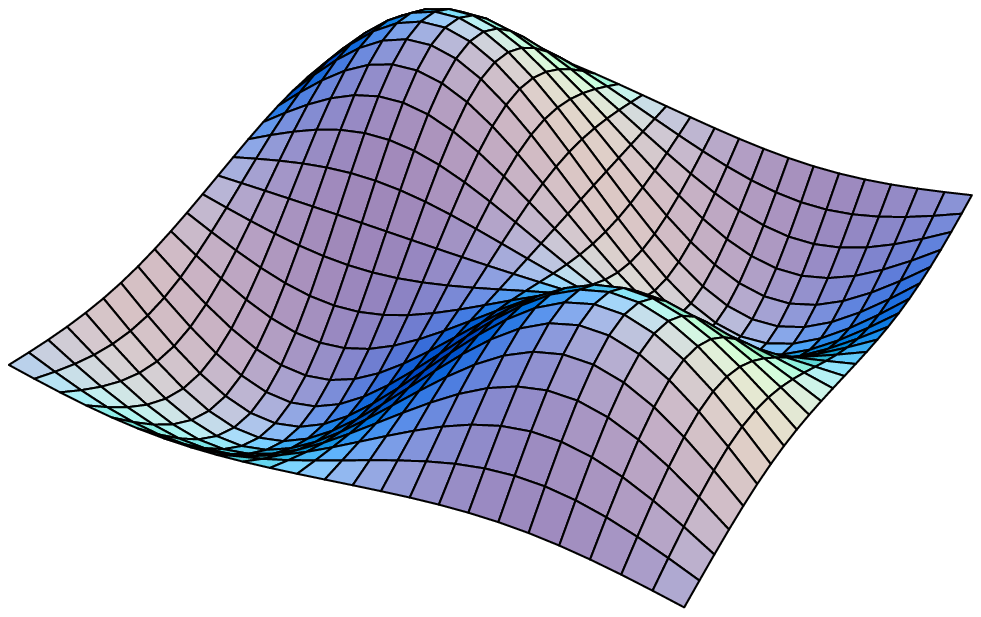}
\caption{The result of adiabatic cooling, starting at finite temperature
with two charge 1 calorons in the ``crossed" configuration (see the text)
on a $16^4$ lattice with $\xi=4$ after 1000 cooling sweeps with $\veps=0$.
The finite temperature solution is presented in the top two figures. The 
left plot shows the action density integrated over time, the right plot 
the Polyakov loop in the time direction, both in the $y$-$z$ plane at $x=8$ 
(where all constituents lie by construction). We changed $\xi$ through 
$2\sqrt2$, 2, $\sqrt2$ to reach 1, at each of these applying 1000 cooling 
sweeps with $\veps=0$. The result at $\xi=1$ is given in the bottom two 
figures (showing the same quantities as above).}
\label{fig:fig7}
\end{figure}
The starting configuration is shown in the top row of Fig.~\ref{fig:fig7}, 
based on improved ($\veps=0$) cooling to stabilize to an exact solution 
(the finite volume modification of the analytic solution for the appropriate 
``crossed" configuration~\cite{Us}). Starting from this optimally localized 
configuration we apply the adiabatic cooling method and find that the lumps 
grow and inevitably overlap, giving rise to extended structures, see the bottom
row of Fig.~\ref{fig:fig7}. It is important to note we used $\veps=0$ cooling 
so as to avoid the cooling to change the moduli of the self-dual solution 
(other than by the changing temperature). This was also to prevent being 
``attracted" to the constant curvature configuration, although we observed 
that actually the result in Fig.~\ref{fig:fig7} forms a local minimum for 
the over-improved action. For the action density we show in this figure 
only the density integrated along the time direction, in the plane going 
through the constituents. Initially, at finite temperature the configuration 
is static. After the adiabatic cooling this is no longer the case. Among any
of the two dimensional slices to be considered no localized structures 
were found. It would not serve a purpose to illustrate this here in further 
detail, but the structures we found look quite similar in nature to those 
shown in Ref.~\cite{DefC}. We also looked at the periodic and anti-periodic 
(w.r.t.~``time") zero-modes to verify the absence of localized structures.

Therefore, well-localized lumps at zero temperature for these low-charge 
self-dual backgrounds can only be found as instantons, even though it is 
clear that these are build from constituents of fractional topological charge. 
On the basis of the caloron solutions at finite temperature, a good guess 
is that the size $\rho$ of the instanton is related to the distance $d$ 
between its constituents as $\hat d=\pi\hat\rho^2$, where $\hat d=d/L$ and 
$\hat\rho=\rho/L$. This indeed provides a good explanation for all the 
features we found, also when using twisted boundary conditions to be 
discussed in the next section.

\section{Twisted Boundary Conditions}\label{sec:tbc}

In this section we consider (minimally) twisted boundary conditions, such 
that it does not affect the topological charge sectors. This is called 
orthogonal twist, and it is best described by the fact that doubling the 
box in just one of the coordinate directions removes the twist (and of 
course doubles the topological charge). The main reason for considering
these boundary conditions is to avoid the obstruction for exact charge 1 
solutions~\cite{BMT}. This way we can be assured that the cooling only 
affects the distance between the constituents~\cite{MTAP}. 

\begin{figure}[htb]
\vskip5.2cm
%\special{psfile=fig8a.eps voffset=-5 hoffset=190 vscale=60.0 hscale=60.0}
\includegraphics{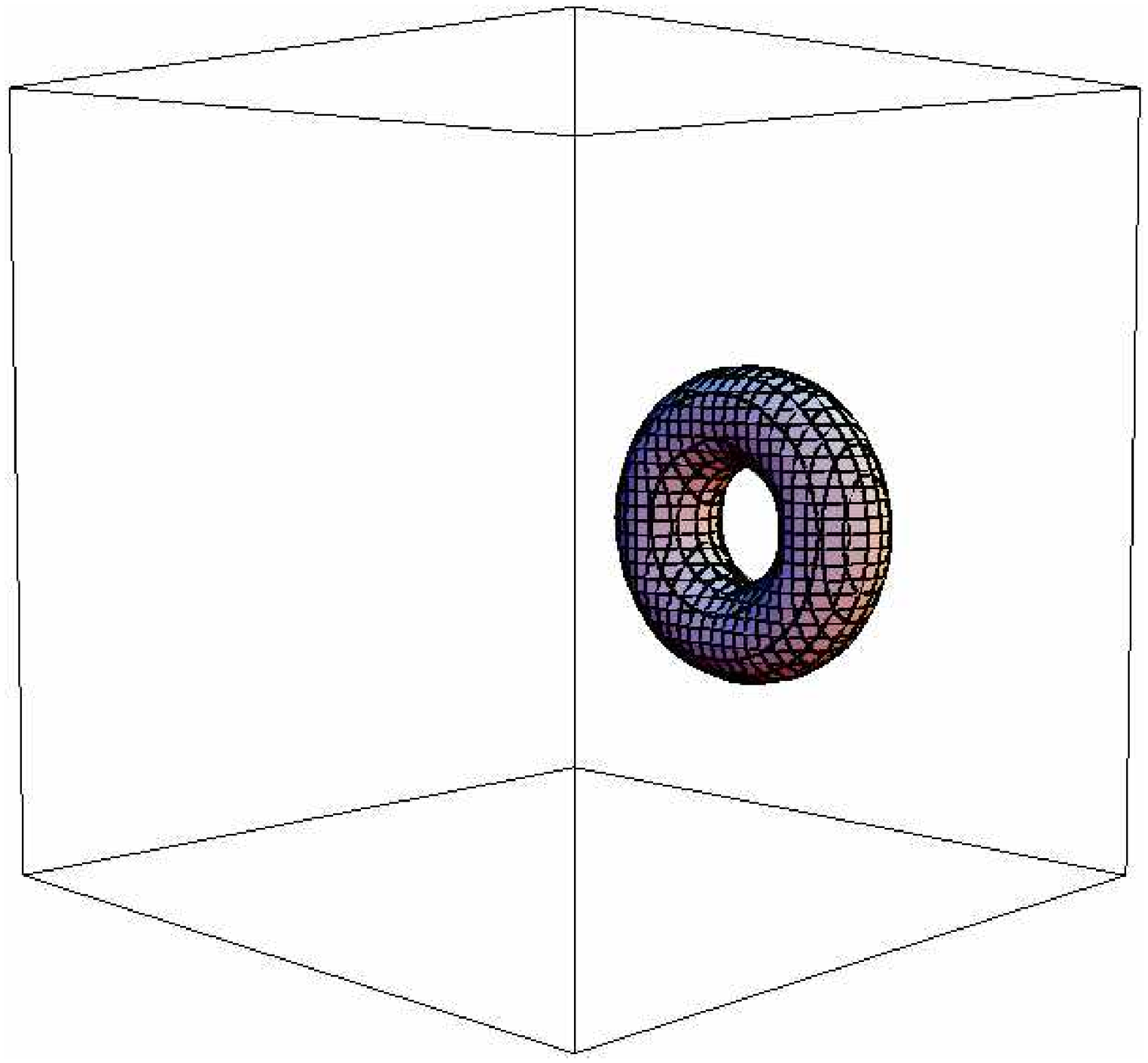}
\includegraphics{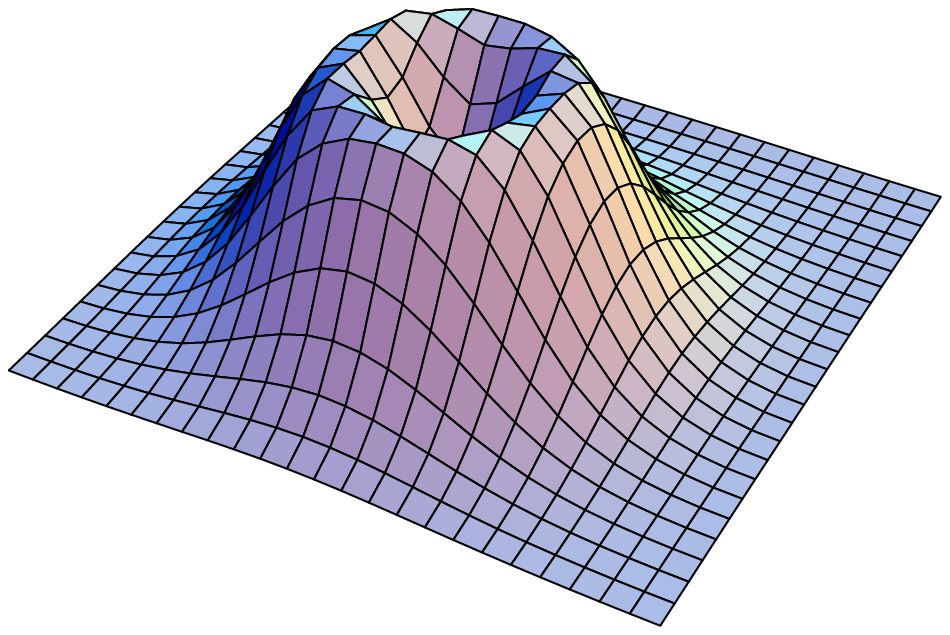}
\caption{A charge 1 configuration on a $4\times 16^3$ lattice with twisted 
boundary conditions in the time direction, $\vec k=(1,0,0)$. It was obtained
from a configuration, first cooled down with $\veps=1$ to slightly above the 
one-instanton action, applying $10^4$ cooling sweeps with $\veps=-10$ (500 
cooling sweeps with $\veps=0$ were finally applied to bring it close to 
the continuum). The constituents had been pushed so far apart that one of 
them effectively changed its electric and magnetic charge. The doughnut 
characteristic for two coinciding magnetic monopoles, with its symmetry axis 
along $\vec k$, gives the ultimate fixed-point under over-improved cooling. 
The right plot shows an action density contour plot of the doughnut, on the 
left the action density is plotted over the $y$-$z$ plane slicing the doughnut 
in two. Gluing two of these boxes along the $\vec k$ direction can be compared 
to Fig.~\ref{fig:fig5}.}\label{fig:fig8}
\end{figure}

First we consider the case of finite temperature with twist $\vec k$ in the
time direction (i.e. a center flux of $k_j$ units in each of the $0$-$j$ 
planes), performing many more cooling sweeps (tens of thousands) than were 
considered in Ref.~\cite{MTAP}. One would expect that the fixed point under 
over-improved cooling would be two constituents maximally separated, i.e. 
by half the size of the box. Somewhat surprisingly this turned out not to 
be the case and the constituents started to get closer together again 
``across the boundary" with further over-improved cooling. This seems in 
contradiction with the fact that constituents of opposite magnetic charge 
repel each other under over-improved cooling. Ultimately we reached the 
situation where the two constituents actually met and formed a doughnut 
structure characteristic of two coinciding magnetic monopoles of the same 
charge, as shown in Fig.~\ref{fig:fig8} for $\vec k=(1,0,0)$. Indeed, 
like-charge constituents attract as we have seen in the previous section. 
We can only conclude that in the process of separating the constituents 
the magnetic charge of one must have changed relative to the other 
constituent. Recalling that the sign of the magnetic charge is correlated 
to the sign of the Polyakov loop observable, this behaviour is related to 
the fact that the Polyakov loop is anti-periodic in certain directions.

That the twisted boundary conditions interfere with the notion of magnetic 
and electric charge is also seen from the charge 1/2 instanton at finite 
temperature, which for all practical purposes behaves as a single
constituent monopole as demonstrated in Ref.~\cite{Syn}. At first sight 
this seems impossible, because a net electric or magnetic charge cannot 
occur in a box with periodic boundary conditions. In this sense twisted 
boundary conditions play a similar role as C-periodic boundary 
conditions introduced in Ref.~\cite{CPer}. 

We may construct a finite volume caloron by putting two boxes with topological 
charge 1/2 next to each other, such that the twist in the space direction 
cancels. This is precisely where the constituents are maximally separated 
and where their charge is ambiguous. When this caloron configuration is 
approached from a localized instanton, whose constituents are pushed apart 
by over-improved cooling, they would be assigned opposite magnetic charge. 
On the other hand, when approached from the doughnut configuration (achieved 
by ordinary cooling as the reverse of over-improved cooling), they would be 
assigned equal magnetic charge. 

\begin{figure}[htb]
\vskip5cm
\includegraphics{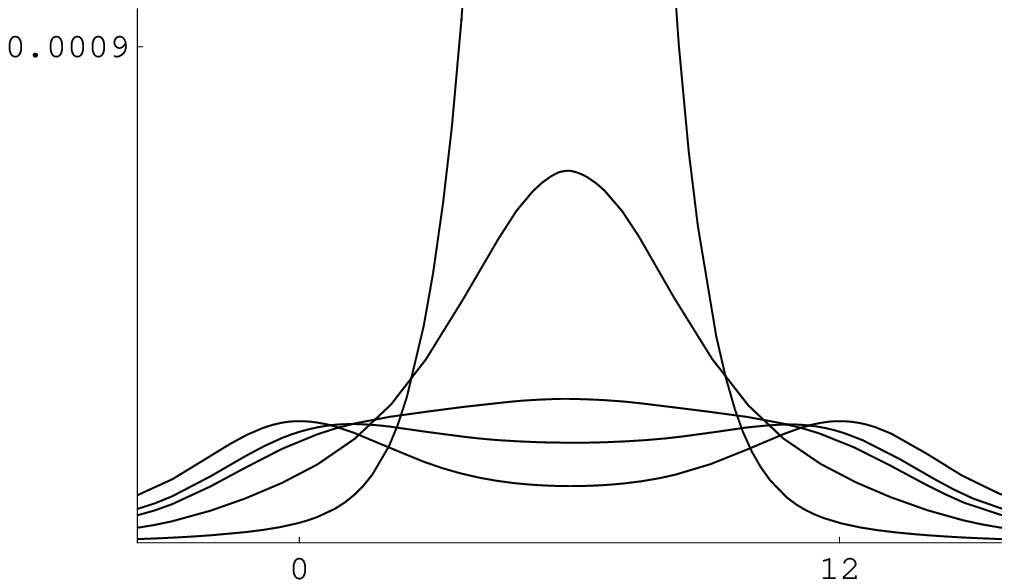}
\includegraphics{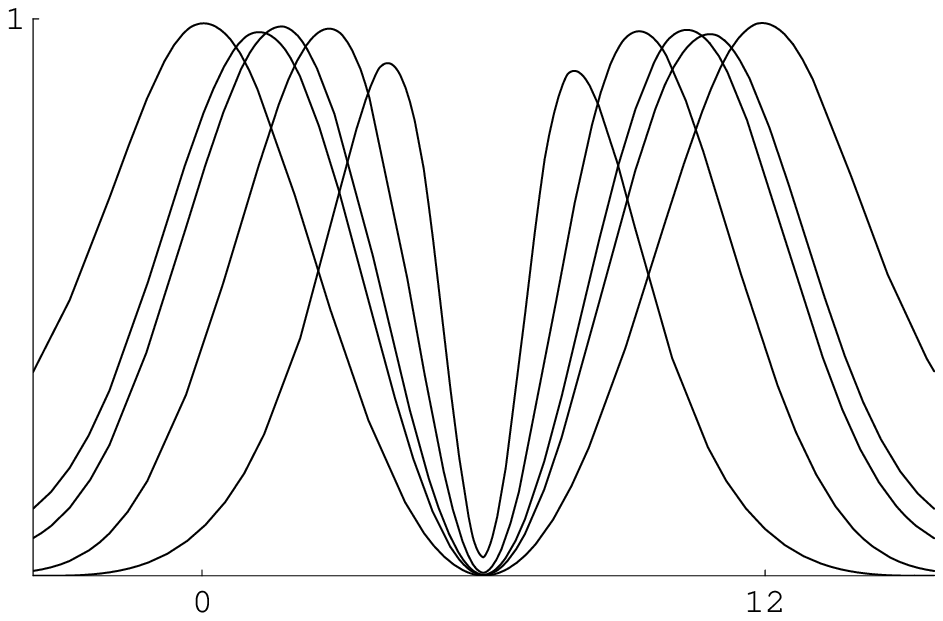}
\caption{Results obtained with cooling on a $12^4$ lattice with twist
in the time direction given by $\vec k=(1,1,1)$, starting from a random
configuration. We first applied 1000 $\veps=1$ cooling sweeps to go down to
slightly above the one-instanton action. We plot the action density (left)
and the {\em square} of half the trace of the Polyakov loop (right) along
the line connecting the two constituents, every time adjusting to the 
continuum by 500 $\veps=0$ additional cooling sweeps. Each curve, with 
constituents pushed further apart, is obtained after ($1000,2000,2000,44000$)
additional $\veps=-10$ cooling sweeps.}\label{fig:fig9}
\end{figure}

We also studied these twisted boundary conditions for the symmetric box. Here 
it is of course a matter of convention what we call the time direction. With 
respect to that arbitrary direction we took for the twist $\vec k=(1,1,1)$. 
We started from a localized charge one instanton. Over-improved cooling will 
automatically start to separate the constituents. We follow this to the point 
where the constituents are close to maximally separated as allowed by the box, 
in which case we can actually distinguish two lumps in the action density, 
which we plot in Fig.~\ref{fig:fig9} along the line connecting the two 
constituents (based on an interpolation of the lattice data). 
We find that only for the maximal separation two individual lumps are visible
in the action density, but that this requires fine-tuning of the placing of 
these constituents; even then the lumps are as big as (half) the volume and 
cannot be considered localized. We have also plotted the {\em square} (due to 
the anti-periodicity) of half the trace of the Polyakov loop which allows us 
to clearly localize the separated lumps.

\section{Cooling Histories}\label{sec:plat}

In this section we will show how cooling histories can be used to establish 
the existence of fractionally charged lumps. It is based on analyzing the 
action and topological charge as measured at plateaux, of which there can 
be many within a given cooling run. A plateau is defined as the point of 
inflection for the action as a function of the cooling sweeps (i.e. when 
the decrease per step becomes minimal)~\cite{Ilg2}.
On the one hand, localized lumps of fractional topological charge can 
annihilate with another lump with the opposite fractional topological charge. 
This would change the overall action (always measured in units of the 1 
instanton action) by twice the value of this topological charge (ranging 
between 0 and 2 depending on the holonomy, but typically around 1), and 
leave the topological charge unchanged. This can thus be easily distinguished 
from the annihilation of an instanton and anti-instanton, for which the 
action always changes by two units. On the other hand two of these lumps with 
the same sign for their fractional topological charge, but opposite magnetic 
and electric charge, can come together and form a localized instanton, which 
subsequently shrinks under cooling (in this section we always use $\veps=1$) 
and then falls through the lattice. In this case both the topological charge 
and the action changes by one unit. 

\begin{figure}[htb]
\vskip7.8cm
\includegraphics{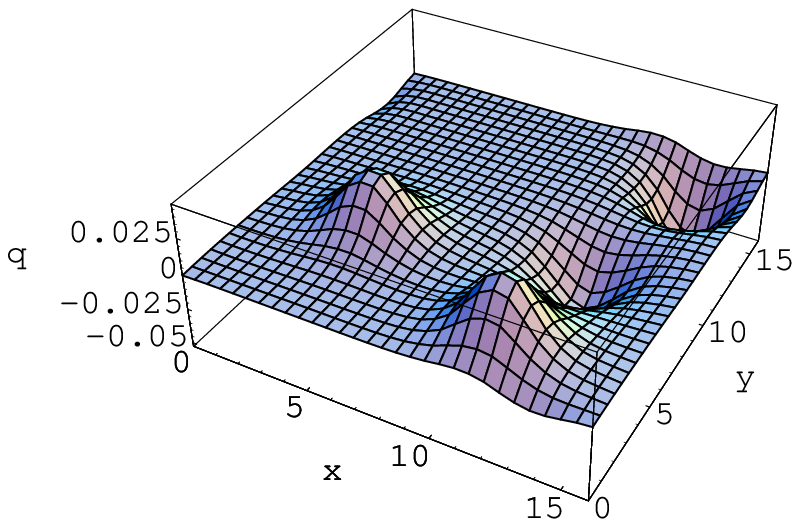}
\includegraphics{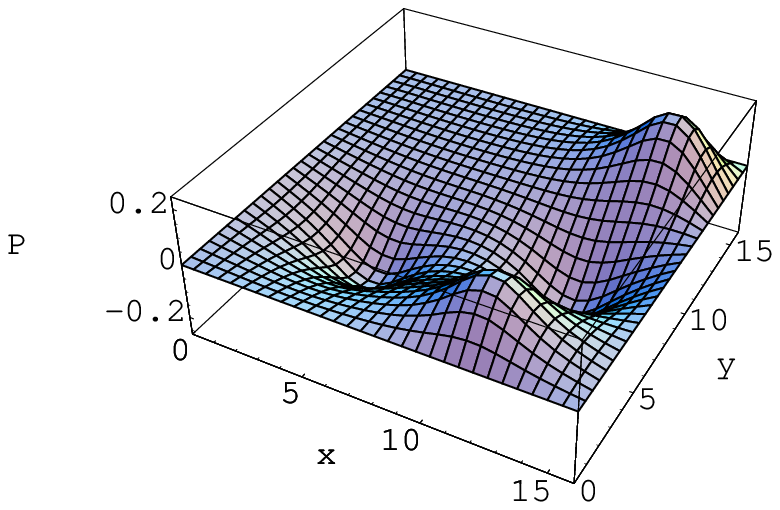}
\includegraphics{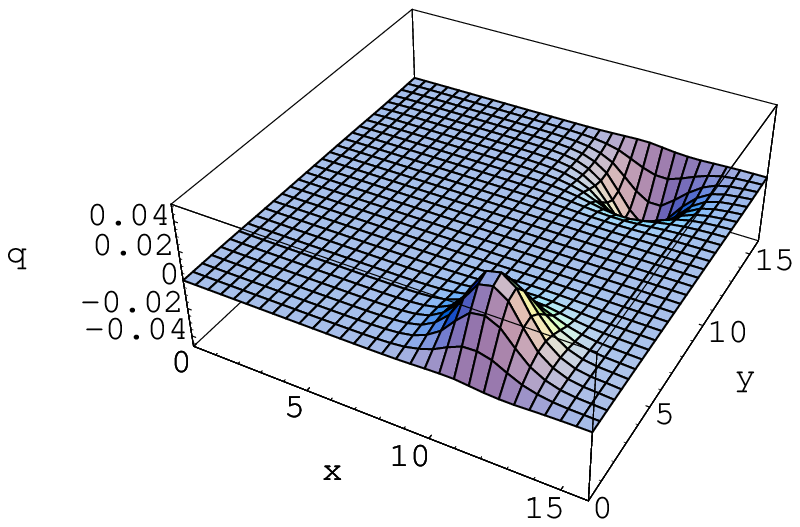}
\includegraphics{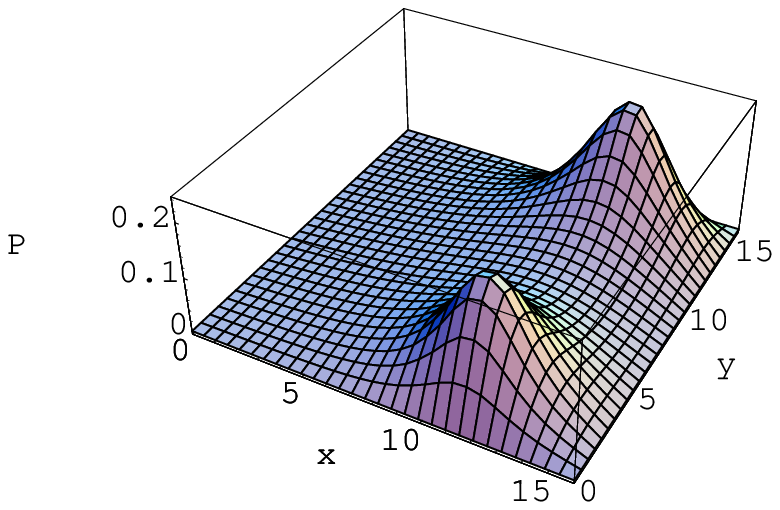}
\caption{Example for annihilation of constituents with opposite fractional
topological charge on a $4\times16^3$ lattice generated from a configuration
just below $T_c$ with ordinary cooling. Shown are, for two consecutive plateaux,
the topological charge density (left) and the Polyakov loop (right) in the
$x$-$y$ plane, averaged over $z$ (and $t$, though the configurations are
nearly static). The annihilation is between constituents with opposite
magnetic, but equal electric charge and equal Polyakov loop.
}\label{fig:fig10}
\end{figure}

\begin{figure}[htb]
\vskip1.3cm\hskip1.7cm\hbox{Im$\,\lambda$}\vskip8.25cm
%\special{psfile=fig11a.eps voffset=32 hoffset=153 vscale=70.0 hscale=70.0}
\includegraphics{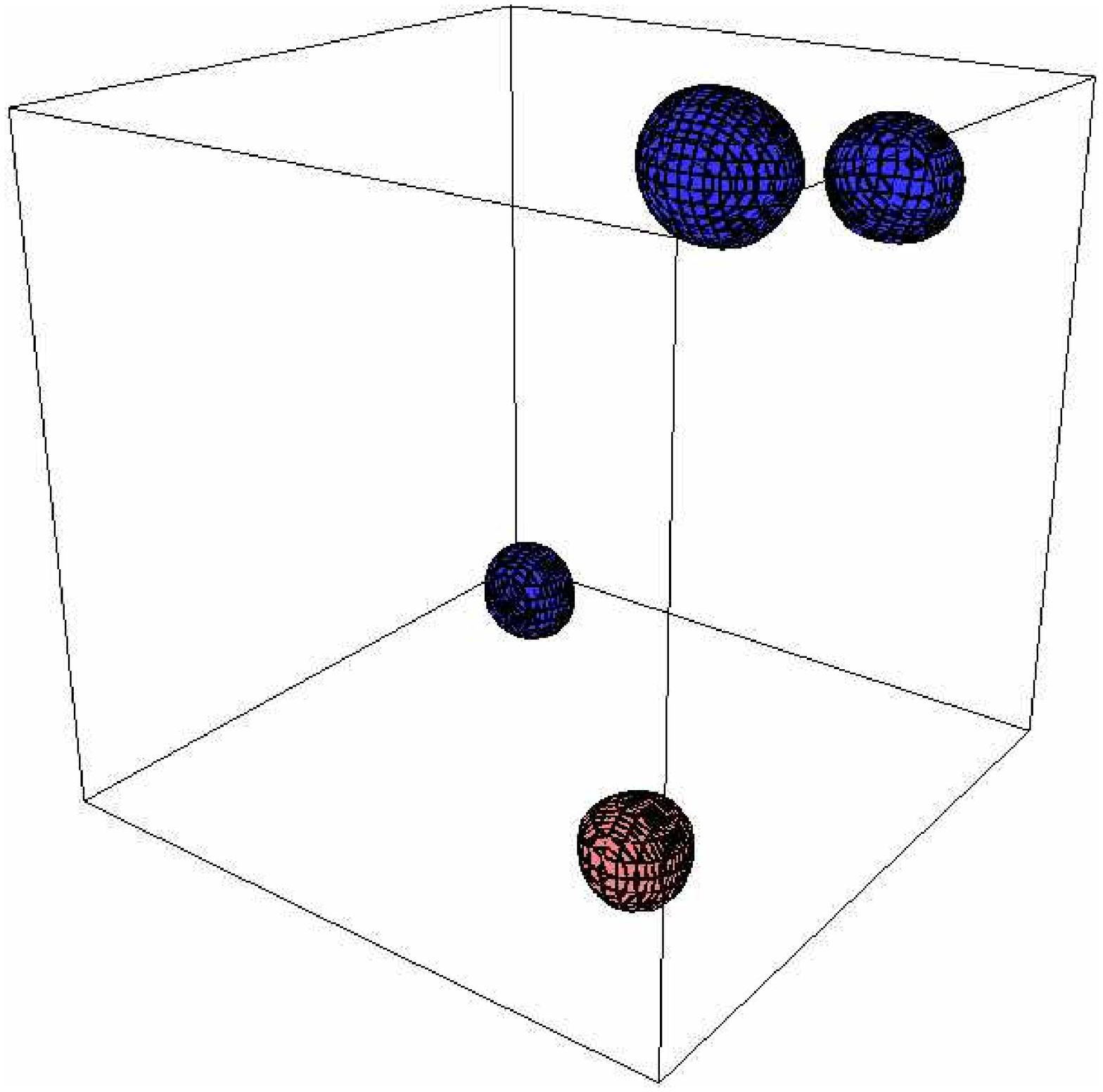}
\includegraphics{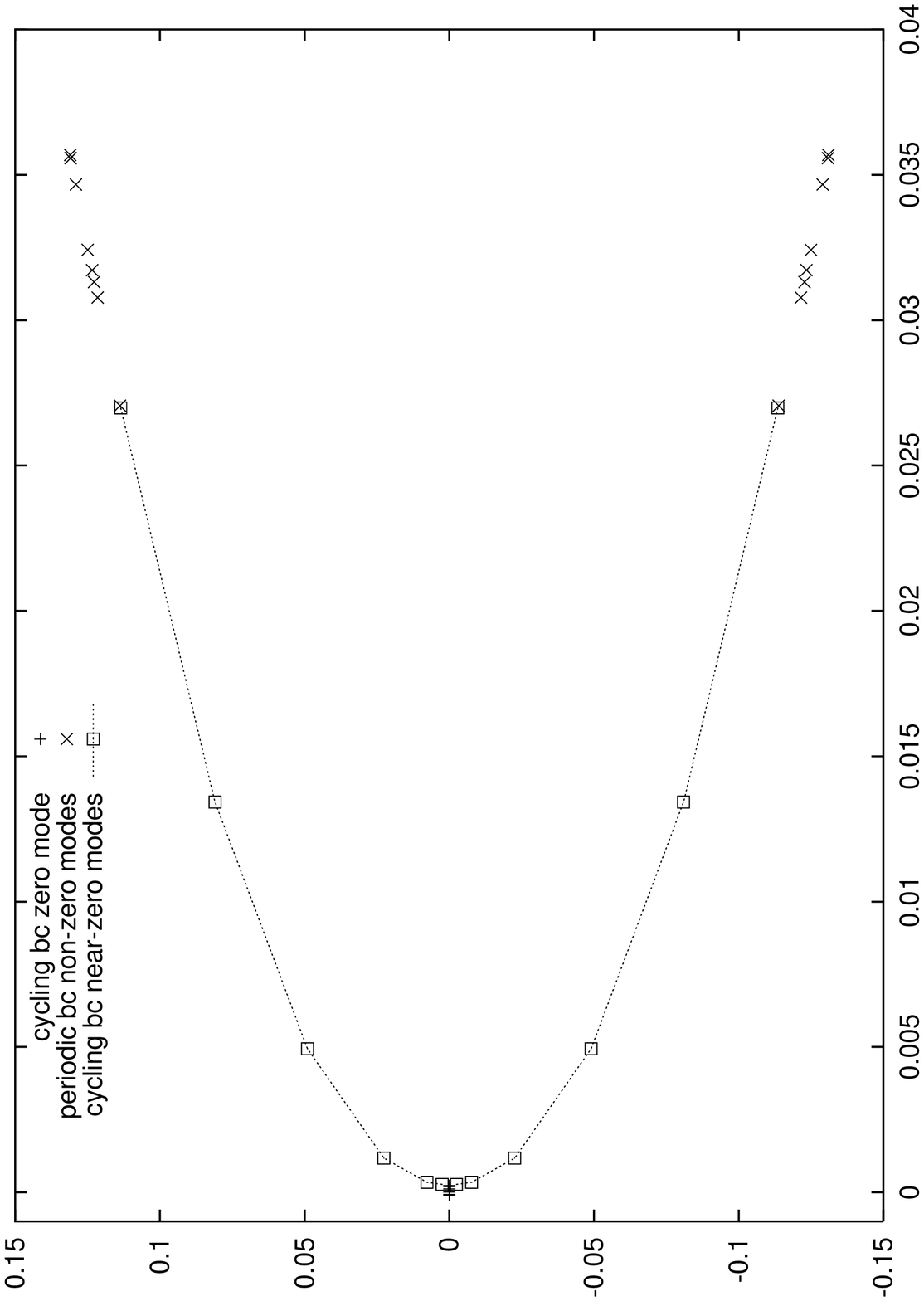}
\vskip-6.95cm\hskip12cm\hbox{\tiny $P\!=\!1\quad~P\!=\!-1$}\vskip2.0cm
\hskip10.4cm\hbox{\tiny $P=-1$}\vskip1.3cm\hskip10.9cm\hbox{\tiny $P=-1$}
\vskip1.45cm\hskip10.5cm\hbox{Re$\,\lambda$}\vskip0.4cm
\caption{Example of a plateau configuration on a $4\times16^3$ lattice (with $S
=1.92$ units and $\half\Tr\pl=-0.22$) before annihilation of the two bottom 
constituents in the inset (light (red) and dark (blue) shading distinguishes 
positive from negative topological charge). The Polyakov loop at the center 
of each of the constituents is indicated by $P=\pm1$. The crosses give the 
low-lying eigenvalues $\lambda$ of the clover-improved Wilson-Dirac operator 
with periodic boundary conditions in time. The two curves trace the two near 
zero-modes from their value with anti-periodic (left) boundary conditions to 
the periodic (right) case (the squares correspond from left to right with 
$n$=$5,4,3,2,1,0$, defining the phase $\exp(2\pi i n/10)$ for the fermion 
boundary conditions).}\label{fig:fig11}
\end{figure}

Examples for the annihilation of constituents of opposite topological charge 
have already been discussed in Ref.~\cite{Ilg2,IMMV}. Here we present two more 
interesting cases. Fig.~\ref{fig:fig10} shows how such pairs of constituents 
with opposite fractional topological charge (bottom row) typically come from a 
caloron and an anti-caloron (top row), after annihilation of the complementary 
pair. In Fig.~\ref{fig:fig11} we present another example in the sector with 
topological charge $-1$, consisting of one close pair of constituents that 
forms an anti-caloron, and a pair of well-separated constituents with opposite 
fractional topological charge. As an inset we show a surface with constant 
topological charge density. 
The colour distinguishes between the three constituents with 
negative and one constituent with positive topological charge. At the next 
plateau (reached after 236 additional cooling sweeps, not shown), the bottom 
pair has annihilated and the constituents of the anti-caloron came closer. 
Fig.~\ref{fig:fig11} also shows the fermion spectrum for the first plateau.
Crosses indicate the low-lying eigenvalues for the Wilson-Dirac operator for 
fermion boundary conditions that are periodic in time. The three lowest 
eigenvalues are traced as a function of the phase of the fermion boundary 
conditions, moving from anti-periodic (left) to periodic (right) boundary 
conditions. The imaginary part of the exact zero-mode stays zero as it 
should, whereas near zero-modes move away from zero.

The (exact positive chirality) periodic zero-mode is localized to the 
constituent with $P=1$ for both plateaux. With anti-periodic boundary 
conditions the (exact positive chirality) zero-mode for the second plateau 
``sees" the one remaining constituent with $P\!=\!-1$. However, for the first 
plateau there are altogether three such (exact and near) zero-modes. One 
of these is the exact positive chirality zero-mode guaranteed by the index 
theorem. It is concentrated only on the two constituents with negative 
topological charge and $P\!=\!-1$. The other two near zero-modes are 
localized on all three constituents with $P\!=\!-1$. But projection on 
the (nearly equal) negative and positive chirality components of the near 
zero-modes will localize to the appropriate constituent(s) with positive 
and negative topological charge. 

\begin{figure}[htb]
\vskip3.8cm\hskip7.6cm$({\bf s})$\vskip3.8cm
\includegraphics{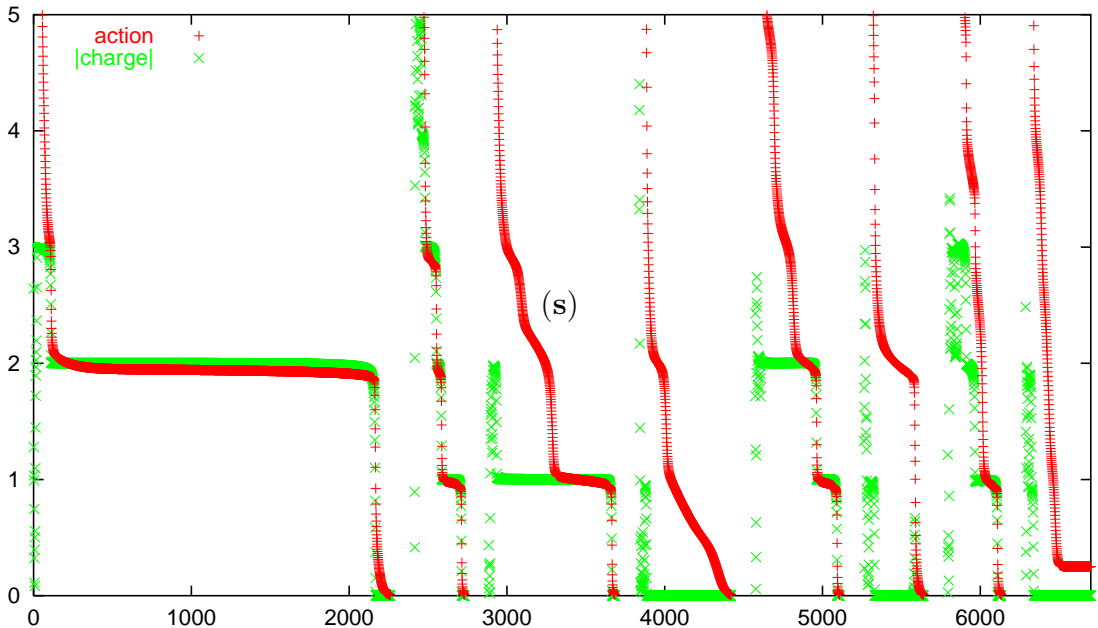}
\caption{Sample of 8 cooling histories on a $4\times 16^3$ lattice
at $4/g^2=2.2$ ($T \approx 0.8 T_c$). Pluses give the Wilson action 
and crosses the absolute value of the (order $a^2$ improved clover 
averaged) topological charge. The curve ``s" is an example discussed
in the text.}\label{fig:fig12}
\end{figure}

\begin{figure}[htb]
\vskip8cm
\includegraphics{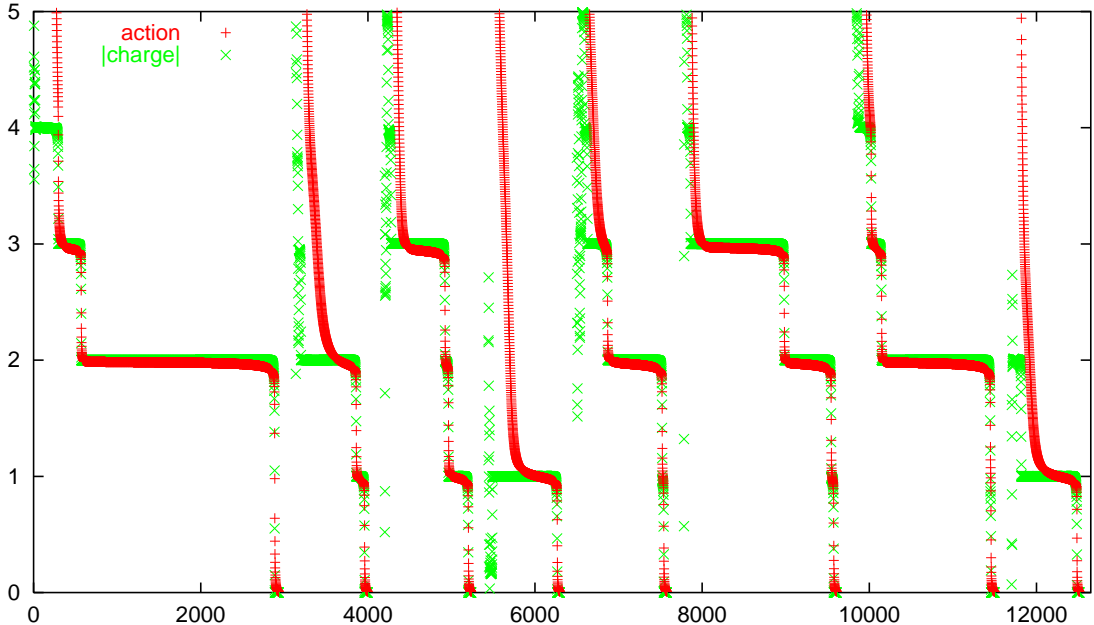}
\caption{Sample of 8 cooling histories on a $16^4$ lattice at $4/g^2=2.3$ 
($T\approx 0.25 T_c$). The symbols are as discussed in the caption of 
Fig.~\ref{fig:fig12}.}\label{fig:fig13}
\end{figure}

We ran Monte Carlo on a $4\times 16^3$ lattice at $4/g^2=2.2$, extracting 50 
configurations equilibrated at $T \approx 0.8 T_c$. A sample of 8 cooling 
histories is shown in Fig.~\ref{fig:fig12}. The crosses give the (order $a^2$ 
improved clover averaged) topological charge, and the pluses the Wilson action 
which was used for the cooling. Much can be read off from this figure. The 
definition of a plateau is the point of inflection, which is also where the 
change in the action is slowed down, as is reflected in the greater density 
of symbols. The ``snake-like" behaviour in the action curves (an example is
indicated by ``s" in the figure), along which the topological charge remains 
constant, represents in many cases examples where constituent annihilation 
takes place. This is so, because the difference in action between the 
consecutive plateaux (i.e.~``bends") is closer to one, rather than to two 
instanton units.  This can be contrasted with the behaviour at low temperature 
($T\approx 0.25 T_c$) with a sample of cooling histories presented in 
Fig.~\ref{fig:fig13}, obtained on a $16^4$ lattice at $4/g^2=2.3$. 

A similar study on a $4\times16^3$ lattice at $4/g^2=2.4$ for the deconfined 
phase was performed as well. In the relatively rare cases that a plateau is 
seen, there is no sign of constituent annihilation. This agrees with our 
expectations, since at trivial holonomy only one of the constituent monopoles 
is massive, capturing all of the action. We do find at finite temperature some
plateaux {\em below} the one-instanton action, which are by now well understood 
as (approximate) constant magnetic field configurations~\cite{IMMV,IMMP}. In 
the confined phase these can be stable depending on the precise value of the 
holonomy, and the last cooling run shown in Fig.~\ref{fig:fig12} provides a 
clear example. The minimal value for a $4\times 16^3$ lattice is one quarter 
of the instanton action, but we found also values twice and three times that 
big.\footnote{In the terminology of Ref.~\cite{IMMP}, the allowed values of 
the action for these constant curvature solutions is $|\vec m|^2/16$ units, 
where $\vec m$ is the magnetic flux, whose components are even integers due 
to the periodic boundary conditions. Provided $|\vec m|< N_s\sqrt\pi/N_t$, 
there is a range of values of the holonomy for which these are (marginally) 
stable, as can be shown from a straightforward generalization of the argument 
given for $\vec m=(0,0,2)$ in Ref.~\cite{IMMP}.}

\begin{figure}[htb]
\vskip10.2cm
\includegraphics{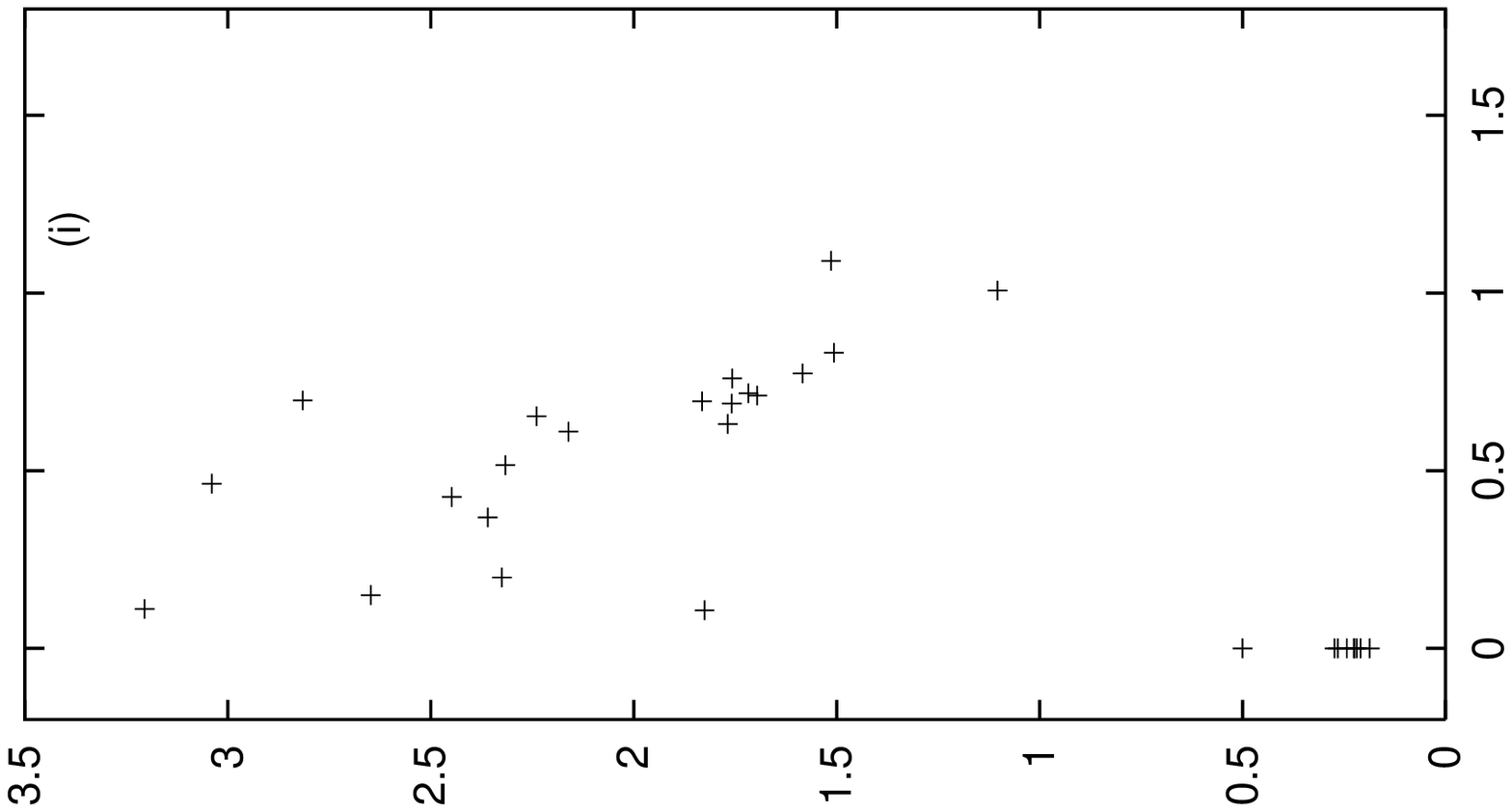}
\includegraphics{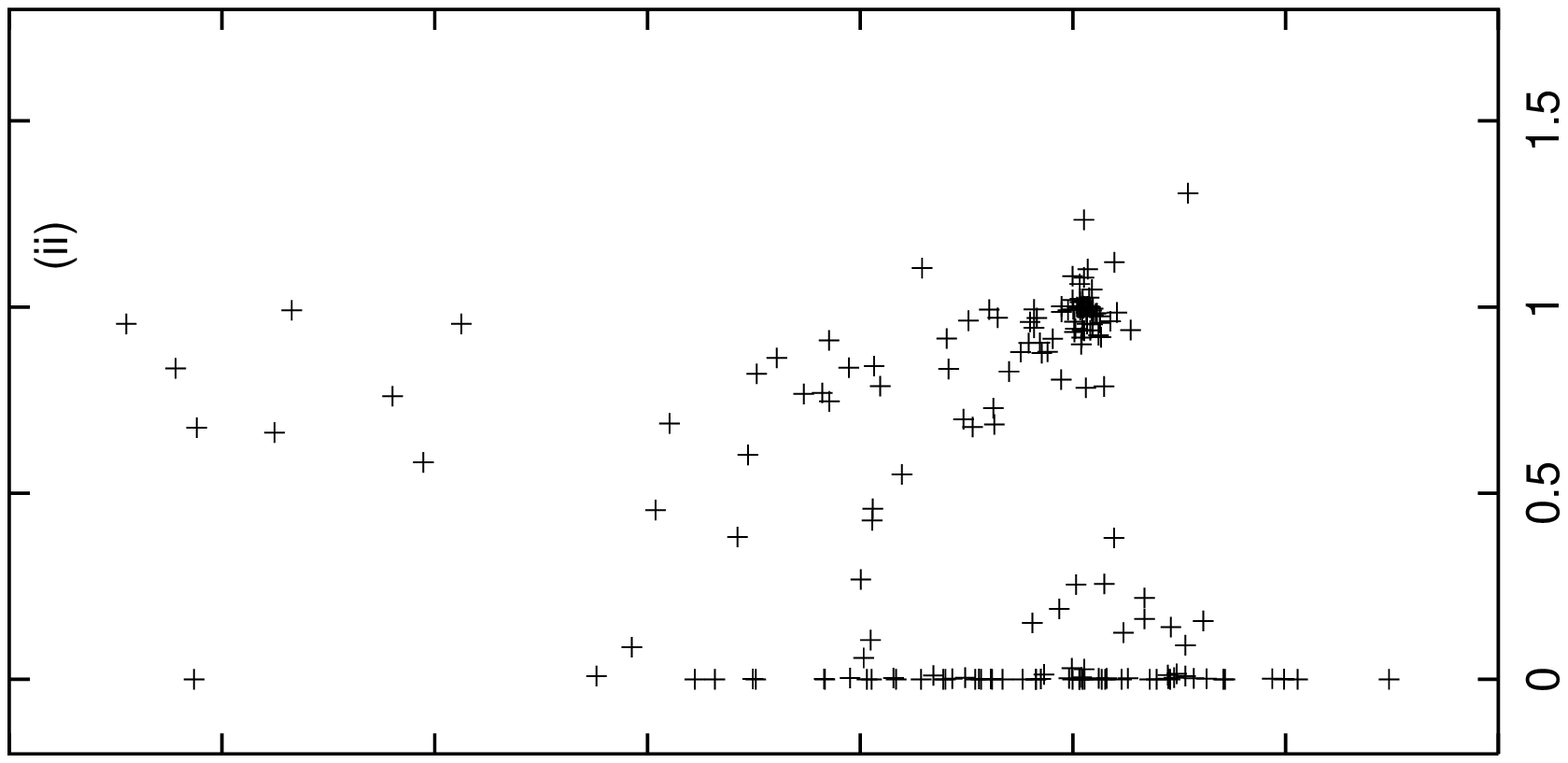}
\includegraphics{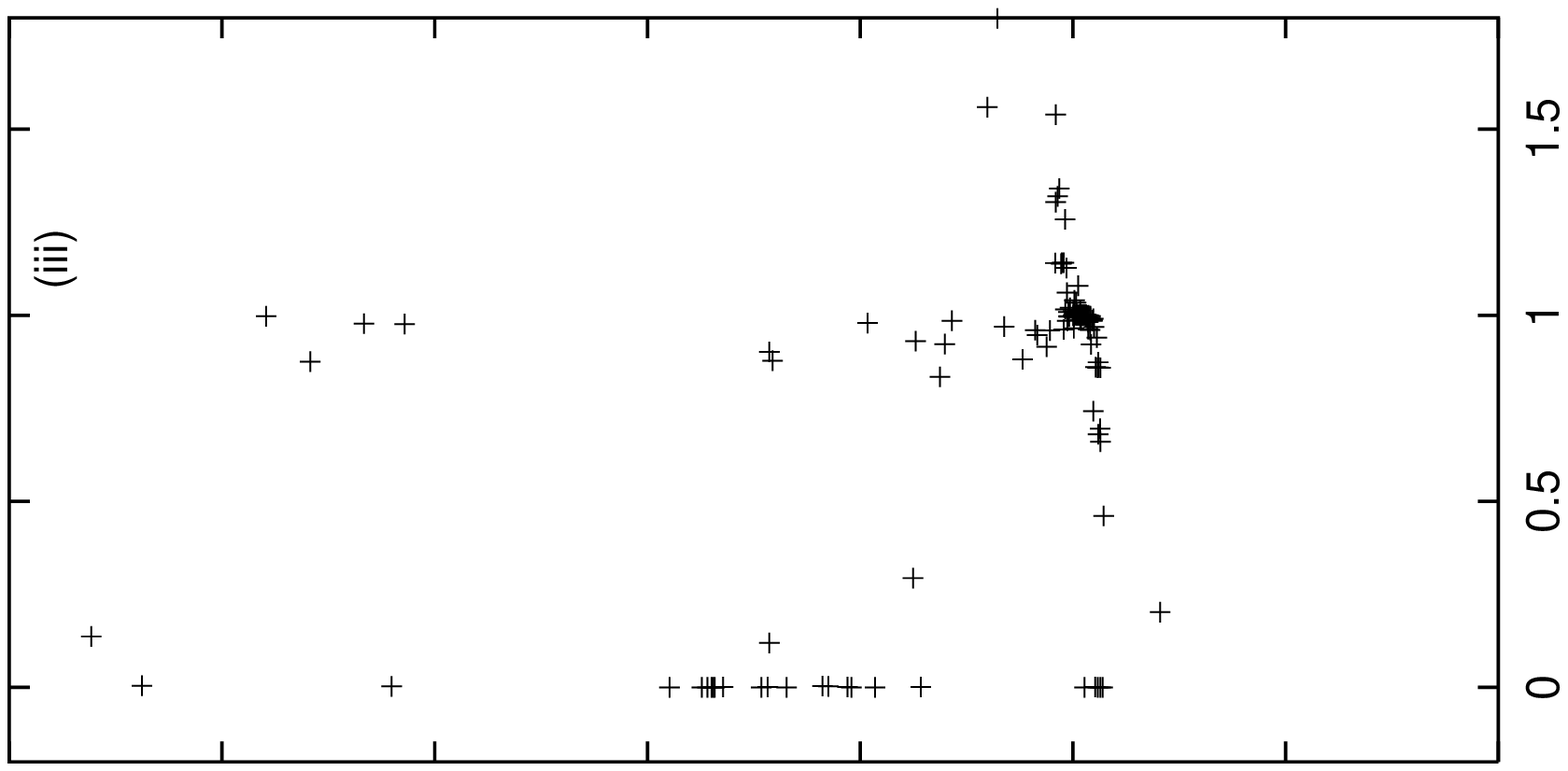}
\caption{Scatter plots of $|\Delta Q|$ (horizontally) versus $\Delta S$ 
(vertically) for (i) a $4\times 16^3$ lattice at $4/g^2=2.4$ ($T\approx 1.2
T_c$), (ii) a $4\times 16^3$ lattice at $4/g^2=2.2$ ($T \approx 0.8 T_c$), 
note the significant clustering around $\Delta S=1$, $|\Delta Q|=0$ 
characteristic of constituent annihilations, and (iii) a $16^4$ lattice 
at $4/g^2=2.3$ ($T\approx 0.25 T_c$). Each case is based on 50
configurations.}\label{fig:fig14}
\end{figure}

We summarize our findings for the plateau analysis in a scatter plot for the 
decrease in action ($\Delta S$) between two plateaux, versus the absolute 
value of the change in the topological charge ($|\Delta Q|$), see 
Fig.~\ref{fig:fig14}. 
Most points in the scatter plot are associated to the typical process of 
instanton disappearance, distributed around $\Delta S=|\Delta Q|=1$. The 
clustering of the points around $\Delta S=1$ and $|\Delta Q|=0$, absent 
for zero and high temperatures, is nevertheless clear evidence for the 
annihilation of {\em localized} constituents with opposite fractional 
topological charge. The action of these constituents depends on the 
holonomy; its fluctuations are reflected in the spread of $\Delta S$ 
around 1. 

\section{Summary and Discussion}\label{sec:disc}

We have analysed the constituent nature of instantons, both at finite 
and zero temperature. As a convenient way to describe the instanton 
moduli space, constituents were long ago conjectured to play a role and 
called instanton quarks~\cite{InQu}. Some early realizations in terms of 
instantons with topological charge $1/n$, that can exist with twisted 
boundary conditions~\cite{THoT,ToMa}, were considered in Ref.~\cite{ToMM} 
(singular solutions like merons~\cite{Mer} excluded). Constituents with 
{\em arbitrary} fractional topological charge were realized at finite 
temperature in the background of non-trivial holonomy~\cite{KvB}. When 
well-separated these are described in a precise way by BPS monopole 
configurations~\cite{BPS}.  In the confined phase, where on average the 
holonomy is maximally non-trivial, the topological charge fraction is 
on average $1/n$. 

Here we have investigated in what sense self-dual solutions of higher 
topological charge are made up of constituents and how the latter overlap. 
Furthermore, we have followed what happens to the constituents in self-dual 
solutions when adiabatically lowering the temperature. As deduced from the 
behaviour of the Polyakov loop, constituents remain present despite the fact 
that they cannot reveal themselves as isolated action density lumps. At zero 
temperature these constituents become massless and are obviously not dilute, 
even though instantons can still be seen as the ``hadrons" made out of these 
constituents. The name instanton quarks is therefore quite appropriate, and 
the possibility of confinement described in terms of a high density ensemble 
of these constituents becomes an appealing one~\cite{Zako}. This would in 
some sense be the ``dual" of deconfinement for high density quark matter, 
even though it remains difficult to quantify this point of view.

Our study has its limitations, since we mainly probe self-dual configurations 
through the cooling studies we performed. In earlier phases of the cooling,
annihilations of constituents (and instantons) of opposite topological charge 
do of course take place. We have even used this to deduce the presence of 
constituents from just studying the cooling histories. This, however, only 
works when the constituents are relatively dilute and well-localized. When 
not dilute, any extended and overlapping structures of opposite topological 
charge will be removed under cooling before being able to be identified. 
More suitable for dynamical studies is the use of chiral fermion zero-modes 
to identify topological structures, as was studied extensively in 
Refs.~\cite{GatS,GatP,Gatt}. To identify the constituents one makes use 
of the fact that when well-separated the localization of the zero-modes 
depends strongly on the boundary conditions used for the fermions~\cite{ZM}. 
At finite temperature these findings agree beautifully with the results 
obtained by cooling. We have demonstrated another effect that can be 
explained by well-separated constituents, namely that the number of 
near zero-modes (in a smooth, but non-selfdual background) can depend
on the fermion boundary conditions.

The signature of well-localized zero-modes changing location when 
cycling through the fermion boundary conditions was also found at 
zero-temperature~\cite{GatP}. A good measure for the localization~\cite{GatS} 
is the inverse participation ratio, $I\equiv V\sum_x\rho^2(x)$, where 
$\rho(x)$ is the zero-mode density. The bigger $I$ is, the more localized 
is the zero-mode. This can be contrasted with a constant zero-mode density
for which $I=1$. On average, at finite temperature~\cite{GatS} $I$ is indeed 
considerably larger than at zero temperature~\cite{GatP}, but in the latter 
case values of $I$ as big as 20 or more are still seen to occur for cases 
where zero-modes jump over distances as large as half the size of the volume 
when cycling through the boundary conditions. On the other hand, at low
temperatures the inverse participation ratios never reached values above~2 
for the zero-modes with maximally separated constituents, as part of 
self-dual configurations.

These studies, using zero-modes as a filter for Monte Carlo generated 
configurations, have not yet provided other independent means to 
distinguish whether a zero-mode is associated to a constituent of fractional 
topological charge or to an instanton with integer topological charge. The 
possibility that the zero-modes are localized to instantons (formed from 
closely bound constituents) and jump between well-separated instantons, 
rather than well-separated isolated constituents, was discussed in 
Ref.~\cite{Tsu}. It was found that at finite temperature this is unlikely 
to occur, but it could not be ruled out for zero temperature. The analysis 
assumes the constituents to be relatively dilute and well-localized, neither 
of which seems to be the case at zero temperature. The argument relies on 
the fact that typically there will be many topological lumps of either sign,
when no cooling is applied to the Monte Carlo generated configurations.
For configurations with exactly one negative chirality zero-mode, one 
minimally requires the presence of $n$ instantons and $n-1$ anti-instantons 
in order for the negative chirality zero-mode to be able to visit $n$ 
different locations, as would be the case for an SU($n$) caloron with 
well-separated constituents.

In a random medium of topological lumps the mechanism of localization of the 
zero-modes could very well be similar to Anderson localization~\cite{AndLoc}.
In such a case one perhaps should expect a dependence on the fermion 
boundary conditions, even when constituents remain well hidden inside 
instantons. In the case that instantons form a dense ensemble, this is 
similar to the statement that it is impossible to determine which set 
of instanton quarks form an instanton. Still, the essential fact remains 
that the boundary conditions determine to which {\em type} of instanton 
quark the zero-mode localizes. That some of the zero-modes, if associated 
to fractionally charged lumps, are more localized than we observed in the
studies presented here can have a dynamical origin. Further work will be 
required to understand all this in more detail, but it seems legitimate to 
conclude that constituents are here to stay, and may well play an important 
role in our understanding of confinement.

\section*{Acknowledgements}

We thank Christof Gattringer, Michael M\"uller-Preussker and D\'aniel 
N\'ogr\'adi for many useful discussions. We also thank Tony Gonz\'alez-Arroyo
for discussions at Lattice 2004 and in particular Margarita Garc\'{\i}a 
P\'erez for providing us with the code of the cooling program with twisted 
boundary conditions and for her generous help in using it. We are grateful
to Dirk Peschka for his help with the zero-mode analysis. We thank Christof 
Gattringer and Andreas Sch\"afer for inviting us to their Regensburg workshop 
``The QCD Vacuum from a Lattice Perspective" and its participants for fruitful 
discussions. This paper was finalized there and started during a two-month 
visit of  E.-M.I. and B.M. to Leiden, who gratefully appreciate the hospitality
experienced at the Instituut-Lorentz of Leiden University. This work was 
supported in part by FOM, by RFBR-DFG (Grant 03-02-04016) and by DFG (grant 
Mu 932/2-1). FB was supported by FOM and E.-M.I. by DFG (Forschergruppe 
Lattice Hadron Phenomenology, FOR 465).

\end{document}